\documentclass[amsmath,amssymb,aps,prd,onecolumn,groupedaddress,nofootinbib]{revtex4-2}

\usepackage{bm}
\usepackage{braket}
\usepackage{graphicx}
\usepackage{mathtools}
\usepackage{dcolumn}

\begin{document}

\title{High-energy neutrinos flavour composition as a probe of neutrino magnetic moments}

\author{Artem Popov}
\email{ar.popov@physics.msu.ru}
\affiliation{Department of Theoretical Physics, \ Moscow State University, Moscow 119991, Russia}
\author{Alexander Studenikin}
\email{studenik@srd.sinp.msu.ru}
\affiliation{Department of Theoretical Physics, \ Moscow State University, Moscow 119991, Russia}

\date{\today}
\begin{abstract}
	Neutrino propagation in the Galactic and extragalactic magnetic fields is considered. We extend an approach developed in \cite{Popov:2019nkr} to describe neutrino flavour and spin oscillations using wave packets. The evolution equations for the neutrino wave packets in a uniform and non-uniform magnetic fields are derived. The analytical expressions for neutrino flavour and spin oscillations probabilities accounting for damping due to the wave packet separation are obtained for the case of a uniform magnetic field. It is shown that terms in the flavour oscillations probabilities that depend on the magnetic field strength are characterized by two coherence lengths. One of the coherence lengths coincides with the coherence length for neutrino oscillations in vacuum, while the second one is proportional to the cube of the average neutrino momentum $p_0^3$. The probabilities of flavour and spin oscillations are calculated numerically for neutrino interacting with the non-uniform Galactic magnetic field. It is shown that oscillations on certain frequencies are suppressed on the Galactic scale due to the neutrino wave packets separation. The flavour compositions of high-energy neutrino flux coming from the Galactic centre and ultra-high energy neutrinos from an extragalactic sourse are calculated accounting for neutrino interaction with the magnetic field and decoherence due to the wave packet separation. It is shown that for neutrino magnetic moments $\sim 10^{-13} \mu_B$ and larger these flavour compositions significantly differ from ones predicted by the vacuum neutrino oscillations scenario.
\end{abstract}
\maketitle

\section{Introduction}
\label{sec:introduction}
High-energy neutrino astronomy is one of the most rapidly developing areas of neutrino physics. In 2013 the IceCube collaboration reported the discovery of extraterrestrial high-energy neutrinos \cite{IceCube:2013low}. Subsequent years of data taking by the IceCube and Baikal-GVD collaborations confirmed existence of the isotropic diffuse flux of high-energy neutrinos \cite{IceCube:2015gsk,IceCube:2016tpw,IceCube:diffuse,Baikal-GVD:diffuse}.

Sources of high-energy neutrino emission remain largely unknown. Theoretical models propose numerous source candidates, such as Active Galactic Nuclei, Gamma-ray bursts, supernovae and hypernovae, supenova remnants, binary systems, pulsars, magnetars and others \cite{Halzen:2002pg,Meszaros:2017fcs,Vissani:2011ea,Zhang:2002xv}. Active galactic nuclei, in particular blazars and quasars, are among the most promising sources. Motivated by coincidence in direction and time of IceCube-170922A neutrino alert and a gamma-ray flare from the blazar TXS 0506+056, an analysis of 9.5 years of IceCube data provided 3.5$\sigma$ evidence of excess of high-energy neutrino events with respect to atmospheric neutrino background from direction of TXS 0506+056 \cite{IceCube:2018-blazar}. This was a first ever evidence of a point source of high-energy neutrinos. Later studies greatly improved our knowledge of astrophysical neutrinos sources. The latest point sources search by IceCube \cite{IceCube:2022der} highlights three possible sources of neutrino emission: blazars TXS 0506+056 and PKS 1424+240, and Seyfert II galaxy NGC 1068, where the latter achieves 4.2$\sigma$ significance. Recent results by Baikal-GVD present additional evidence of neutrino emission from TXS 0506+056, as well as from other astrophysical sources \cite{Baikal-GVD:2022fmn}. Plavin \emph{et al.} argue \cite{Plavin:2020emb,Plavin:2020mkf,Plavin:2022oyy} that high-energy neutrinos may originate from cores of radio blazars. Neutrino emission from the Galactic plane was identified by IceCube with 4.5$\sigma$ significance.
After completion of Baikal-GVD \cite{Baikal-GVD:2020irv,Avrorin:2022lyk} and KM3NeT \cite{KM3Net:2016zxf}, as well as future neutrino telescopes, such as IceCube-Gen2 \cite{IceCube-Gen2:2020qha} and P-ONE \cite{P-ONE:2020ljt}, sensitivity to high-energy neutrinos sources will increase greatly.

Sources of high-energy neutrinos are believed to be connected with cosmic rays accelerators. Within the hadronic ($pp$) and photo-hadronic ($p\gamma$) scenarios, it is expected that high-energy protons and nuclei produced inside an astrophysical object interact with the ambient gas and radiation, creating charged and neutral pions. Neutral pions decay $\pi^0 \to \gamma\gamma$ emit high-energy gamma rays, while charged pions produce neutrinos and antineutrinos in the decay chain $\pi^{+(-)} \to \mu^{+(-)} + \nu_\mu (\bar{\nu}_\mu)$ followed by $\mu^{+(-)} \to e^{+(-)} + \bar{\nu}_\mu (\nu_\mu) + \nu_e (\bar{\nu}_e)$ (see \cite{Ahlers:2018fkn}). The flavour ratios of neutrinos produced in such scenarios follows pattern $\nu_e : \nu_\mu :\nu_\tau = 1:2:0$. Note that in this paper we do not distinguish between neutrino and antineutrinos since they are not well discriminated by current detectors. However, in principle discrimination of $\nu_e$ and $\bar{\nu}_e$ is possible via the Glashow resonance \cite{Glashow}. Alternatively to pions decay mechanism, the muon decay, neutron decay and charm decay scenarios predict $0:1:0$, $1:0:0$ and $1:1:0$ flavour ratios respectively \cite{Barger:2014iua}. Some models predict more complicated energy-dependent flavour ratios \cite{Kashti:2005qa}.

Due to the phenomenon of neutrino oscillations flavour  ratio observed by terrestrial neutrino telescopes is different from flavour ratios at the source. The standard picture of neutrino flavour oscillations in vacuum predicts the flavour composition
\begin{equation}
	r_\alpha  = \sum_{\beta} r^0_{\beta} \sum_{i=1}^3 |U_{\alpha i}|^2|U_{\beta i}|^2,
\end{equation}
where $r^0_{\beta}$ are the flavour ratios at the source.

Different models of physics beyond the Standard Model predict that the observed flavour composition significantly differ from one predicted by the standard vacuum oscillations. Among them are neutrino decay scenarios \cite{Beacom:2002vi,Barenboim:2003jm,Baerwald:2012kc,Bustamante:2015waa}, models with quantum decoherence of neutrinos \cite{Gago:2002na}, theories with the Lorentz-violation \cite{Hooper:2005jp}, the pseudo-Dirac neutrino scenario \cite{Esmaili:2012ac,Brdar:2018tce}.
For review of possible BSM effects and capability of their detection in future neutrino telescopes see also \cite{Pakvasa:2008nx,Shoemaker:2015qul,Ahlers:2018mkf}.

Unlike cosmic rays, neutrinos are not deflected by magnetic fields and travel through cosmic space without absorption. Thus, they are considered to be promising messengers carrying information about processes inside astrophysical objects. However, even if neutrinos are considered to be electrically neutral particles, they still can interact with electromagnetic fields (see \cite{Giunti:2014ixa,Giunti:2025} for a review on neutrino electromagnetic interactions). Massive neutrinos may possess electromagnetic properties, in particular nonzero anomalous magnetic moments \cite{Fujikawa:1980yx,Shrock:1982sc}. For the case of Dirac neutrinos diagonal magnetic moments, Standard Model predicts values
\begin{equation}
	\mu_{ii}^D = \frac{3e G_F m_i}{8\sqrt{2}\pi^2} \approx 3.2\times 10^{-19} \left( \frac{m_i}{\text{eV}} \right) \mu_B,
\end{equation}
where $\mu_B$ is the Bohr magneton. A number of theories of physics beyond the Standard Model predict values of neutrino magnetic moments in the range $10^{-12} \div 10^{-17} \mu_B$ (see \cite{Giunti:2014ixa} for a review). The most stringent terrestrial upper bound on neutrino effective magnetic moment is obtained by GEMMA \cite{Beda:2012zz} experiment and is on the level of $2.9 \times 10^{-11} \mu_B$. Solar neutrinos observation by XENONnT provides the upper bound on the level of $6.4\times10^{-12}\mu_B$ \cite{XENONCollaboration:2022kmb}. There are also a variety of astrophysical limits of order of $10^{-12} \mu_B$ \cite{PDG:2022}.

Phenomena of neutrino spin \cite{Cisneros:1970nq} and spin-flavour precession \cite{Schechter:1981hw} arise due to interaction of neutrino magnetic moments with a magnetic field. In \cite{Kurashvili:2017zab,Popov:2019nkr} it was shown that the probabilities of flavour oscillations in a magnetic field are the combinations of oscillations on both vacuum frequencies $\omega^{vac}_{ij} = \Delta m^2_{ij}/2E$ and magnetic frequencies $\omega_i^B = \mu_i B_\perp$. Spin precession induced by neutrino interaction with cosmic magnetic fields can significantly modify the flavour content of neutrino fluxes detected by the terrestrial neutrino telescopes.

A variety of magnetic field strengths is found in different cosmic structures. Magnetic field of our Galaxy is of order of $\mu$G \cite{Jansson:2012pc}. Furthermore, within the galaxy clusters a magnetic field of strength $\sim 1 \mu$G are found \cite{Valee:2002}. For a review on the galactic and extragalactic magnetic fields see \cite{Beck:2008ty,Han:2017,Govoni:2004as,Ryu:2012}.

Neutrino oscillations in the interstellar magnetic fields was studied before in several papers \cite{Kurashvili:2017zab,Enqvist:1998un,Abeyadira:2021xmn,Lichkunov:2022mjf,Alok:2022pdn,Kopp:2022cug}. However, in these papers the probabilities of neutrino oscillations in a magnetic field were calculated within the plane-wave approximation that does not account for possible decoherence effects due to wave packets separation when neutrinos travel long distances.

To examine possible  decoherence effects during neutrinos propagation one has to use the wave packet approach \cite{Giunti:2003ax}. Previously the wave packet formalism was developed for the case of vacuum neutrino oscillations \cite{Nussinov:1976uw,Kayser:1981ye,Kiers:1995zj,Egorov:2019vqv,Naumov:2020yyv}, neutrino oscillations in matter \cite{Peltoniemi:2000nw,Kersten:2015kio} and neutrino collective oscillations \cite{Akhmedov:2017mcc}. In the present paper the wave packet approach is extended to the case of neutrino propagating in a magnetic field.

The paper is organized as follows. In Section \ref{sec:iii} we develop a formalism for calculation of the neutrino flavour and spin oscillations probabilities in a magnetic field accounting for the wave packets separation effects. In Section \ref{sec:iv} we discuss possible flavour compositions of high-energy and ultra-high energy neutrinos propagating in a magnetic field. Finally, Section \ref{sec:conclusion} concludes our paper.

\section{Neutrino oscillations in a magnetic field in the wave packet formalism}
\label{sec:iii}
In this section we extend the approach to the problem of neutrino oscillations in a magnetic field developed in \cite{Popov:2019nkr} to account for the wave packet separation effects. For illustrative purposes, we start with an analytical solution of neutrino wave packet evolution in a uniform magnetic field. We show that two different coherence lengths appear in the transitions probabilities due to interaction with a magnetic field. Next, in Section 4 we numerically solve the evolution equation using a realistic model of the Galactic magnetic field.

\subsection{Neutrino oscillations in a uniform magnetic field}
\label{sec:3.1}
Neutrino evolution in a uniform magnetic field is described by the following system of modified Dirac equations
\begin{equation}\label{Dirac_eq}
	(i\gamma^{\mu}\partial_{\mu} - m_i)\nu_i(x) + \sum_k \mu_{ik}\bm{\Sigma}\bm{B} \nu_k(x) = 0,
\end{equation}
where $i,k=1,2,3$. Here $\bm{\Sigma}= \begin{pmatrix}
	\bm{\sigma} & 0\\
	0 & \bm{\sigma}
\end{pmatrix}$ and $\bm{\sigma}$ is the vector of Pauli matrices. Neutrino magnetic moment matrices are defined by
\begin{equation}\label{mag_moments}
	\mu^D = \begin{pmatrix}
		\mu_{11} & \mu_{12} & \mu_{13} \\
		\mu_{12} & \mu_{22} & \mu_{23} \\
		\mu_{13} & \mu_{23} & \mu_{33}
	\end{pmatrix}, \;\;\;
	\mu^M = \begin{pmatrix}
		0 & i\mu_{12} & i\mu_{13} \\
		-i\mu_{12} & 0 & i\mu_{23} \\
		-i\mu_{13} & -i\mu_{23} & 0
	\end{pmatrix}
\end{equation}
for Dirac and Majorana neutrinos, respectively \cite{Giunti:2014ixa}. Here $\mu_{ik}$ are the magnetic moments in the neutrino mass basis.

In our previous paper \cite{Popov:2021icg}, Eq. (\ref{Dirac_eq}) was considered in the plane wave approximation that does not account for potentially important decoherence effects in neutrino oscillations at long distances. To account for possible wave packet separation effects we transform Eq. (\ref{Dirac_eq}) to the momentum representation.
In the considered case Eq. (\ref{Dirac_eq}) can be rewritten as
\begin{equation}\label{uniform_B_p_space}
	i\partial_t \nu_i(p,t) = [m_i\gamma_0 + \gamma_0\gamma_3 p]\nu_i(p,t) - \sum_k \mu_{ik}\gamma_0\bm{\Sigma}\bm{B} \nu_k(p,t),
\end{equation}
where the Fourier transform of the neutrino wave function is defined by
\begin{equation}
	\nu_i(x,t) = \int \frac{dp}{(2\pi)^{1/2}} e^{ipx} \nu_i(p,t).
\end{equation}
Here we assume that the neutrino momentum $\bm{p}$ is directed along the $\bm{n}_z$ axis.

We assume that the initial neutrino wave function at $t=0$ in laboratory frame is described by the Gaussian wave packet
\begin{eqnarray}
	\nu_i(p,0) &=& f_i(p,p_0) u^{-}_i(p), \\
	f_i(p,p_0) &=& \frac{1}{(2\pi\sigma_p^2)^{1/4}}\exp\left(-\frac{(p-p_0)^2}{4\sigma_p^2}\right),
\end{eqnarray}
where $p_0$ is the average wave packet momentum, $\sigma_p$ is wave packet width and $u^{-}_i$ is a left-handed solution of vacuum Dirac equation. For the sake of simplicity, we consider one-dimensional wave packets. Note that the three-dimensional wave packets without accounting for a magnetic field presence are considered in \cite{Naumov:2020yyv,Naumov:2013uia}.

In what follows, we consider the case when neutrinos have only the diagonal magnetic moments. In \cite{Kurashvili:2017zab} it was shown that the transition magnetic moments affect patterns of neutrino oscillations in the interstellar magnetic fields for neutrino energies $\sim$ 100 EeV and higher. Thus, the transition magnetic moments are irrelevant of neutrinos detected by IceCube, Baikal-GVD and KM3NeT. In particular, oscillations of high-energy Majorana neutrinos in interstellar magnetic fields are described by the vacuum oscillations probabilities. However, transition magnetic moments become relevant for higher neutrino energies and/or higher magnetic field strength and can induce the resonant enhancement of neutrino oscillations \cite{Chukhnova:2020xth}.

In the case of absence of the transition magnetic moments, Eqs. (\ref{Dirac_eq}) decouple and it becomes possible to describe neutrino oscillations in a magnetic field as the solution of three independent equations for the massive neutrino states:
\begin{equation}\label{Dirac_eq_diag_mm}
	i\partial_t \nu_i(p,t) = ( m_i\gamma_0 + \gamma_0\gamma_3 p - \mu_{i}\gamma_0\bm{\Sigma}\bm{B} ) \nu_i(p,t) \equiv H_i(p) \nu_i(p,t),
\end{equation}
where $\mu_i \equiv \mu_{ii}$ are the diagonal magnetic moments of neutrinos. The form of this equation is similar to the one considered in \cite{Popov:2019nkr,Lichkunov:2022mjf} that allows us to solve it using the same method. The solutions of Eq. (\ref{Dirac_eq_diag_mm}) with initial condition

\begin{equation}
	\nu_i^h(p,0) = f_i(p,p_0) u^h_i(p)
\end{equation}
can be represented \cite{Popov:2019nkr,Lichkunov:2022mjf} as a superposition of states with the definite helicity $h' = \pm 1$:
\begin{equation}\label{superposition}
	\nu_i^h(p,t) = \sum_{s,h'} C_{is}^{h h'} e^{-i E_i^s(p) t} f_i(p,p_0) u_i^{h'},
\end{equation}
where the dispersion relation is given by the eigenvalues of $H_i(p)$
\begin{equation}\label{energy}
	E_i^s(p) = \pm\sqrt{m_i^2 + p^2 + \mu_i^2 B^2 - 2s\mu_i\sqrt{m_i^2 B^2 + p^2 B_{\perp}^2 }},
\end{equation}
where $s = \pm 1$. Here we decompose the magnetic field $\bm{B}$ into the transverse $\bm{B}_{\perp}$ and the longitudinal $\bm{B}_{\parallel}$ components with respect to the neutrino momentum. Note that the interaction with a magnetic field can induce transitions between the neutrino helicity states.

The quantum number $s=\pm1$ in (\ref{energy}) enumerates eigenstates of neutrinos in a magnetic fields that are eigenvectors of the spin operator (see \cite{sokolov_ternov})
\begin{equation}\label{spin_oper}
	S_i = \frac{m_i}{\sqrt{m_i^2 B^2 + p^2 B^2_{\perp}}} \left[ \bm{\Sigma}\bm{B} - \frac{i}{m_i}\gamma_0 \gamma_5 [\bm{\Sigma}\times\bm{p}]\bm{B}\right],
\end{equation}
which commutes with the Hamiltonian $H_i(p)$ introduced in (\ref{Dirac_eq_diag_mm}).

For a sufficiently narrow wave packet, one can assume $u_i^h(p) \approx u_i^h(p_0)$. Furthermore, in the ultrarelativistic limit that is obviously justified for high-energy neutrinos, we can neglect terms of order of $m_i/p$ and get
\begin{equation}
	u_i^- \approx \frac{1}{\sqrt{2}} \begin{pmatrix}
		0 \\
		-1 \\
		0 \\
		1
	\end{pmatrix}, \ \ \
	u_i^+ \approx \frac{1}{\sqrt{2}} \begin{pmatrix}
		1 \\
		0 \\
		1 \\
		0
	\end{pmatrix}.
\end{equation}
In this case, as it was shown in \cite{Popov:2019nkr}, the coefficients $C^{hh'}_{i s}$ are given by
\begin{equation}\label{LL}
	C^{LL}_{i s} \approx \frac{1}{2} + \mathcal{O}\Big(\frac{m_i^2}{p^2}\Big), \;\;\; C^{RL}_{i s} \approx - \frac{s}{2} + \mathcal{O}\Big(\frac{m_i^2}{p^2}\Big).
\end{equation}

The dispersion relation (\ref{energy}) can be decomposed near the average momentum $p_0$
\begin{equation}\label{dispersion_decomposition}
	E_i^s(p) = E(p_0) + v_i^s(p_0)(p-p_0) + \mathcal{O}((p-p_0)^2),
\end{equation}
where the neutrino wave packets group velocities are introduced
\begin{equation}\label{group_velocity}
	v_i^s(p_0) = \frac{\partial E_i^s(p)}{\partial p}\Big|_{p=p_0} = \frac{p_0}{E_i^s(p_0)}\left(1-\frac{s\mu_i B^2_{\perp}}{\sqrt{m_i^2 B^2 + p^2_0 B^2_{\perp}}} \right).
\end{equation}

Consider a particular case of the transversal magnetic field $\bm{B} = \bm{B}_\perp$. In this case the dispersion relation (\ref{energy}) takes the form
\begin{equation}
	E_i^s(p) = \sqrt{m_i^2 + p^2} - s\mu_i B_{\perp}
\end{equation}
and the group velocities are given by
\begin{equation}\label{group_vel_vac}
	v_i^s = \dfrac{p}{\sqrt{m_i^2 + p^2}}.
\end{equation}
Eq. (\ref{group_vel_vac}) coincides with the group velocities of neutrinos propagating in vacuum and do not depend on the spin number $s$. Thus, the separation of neutrino wave packets with different spin numbers $s$ is caused by the longitudinal component of the magnetic field $\bm{B}_\parallel$.

For ultrarelativistic neutrinos, assuming that $\mu_i B \ll m_i$ and $B_\parallel \sim B_\perp$, we obtain the following series expansion of the group velocities (\ref{group_velocity})
\begin{equation}\label{group_vel_expansion}
	v_i^s(p_0) = 1 - \frac{m_i^2}{2 p_0^2} - \frac{\mu_i^2 B_{\perp}^2}{2 p_0^2} + \frac{s \mu_i B_{\perp} m_i^2}{p_0^3} \left( \frac{B^2}{B_{\perp}^2} + 3 \right) + \mathcal{O}\left(\frac{m_i^2}{2 p_0^2} + \frac{\mu_i^2 B_{\perp}^2}{2 p_0^2}\right)^2.
\end{equation}

After substituting (\ref{dispersion_decomposition}) into (\ref{superposition}) and performing the Fourier transform back into the coordinate space, one can obtain the following expression for the neutrino wave function:

\begin{equation}\label{wave_fun_x_space}
	\nu_i^h(x,t) = \frac 1 N \sum_{s,h'} C_{is}^{hh'} e^{-iE_i^s(p_0)t + ip_0 x} \exp\Big({-\frac{(x-v_i^st)^2}{4\sigma_x^2}}\Big) u_i^{h'},
\end{equation}
where $h$ and $h'$ are the initial and final neutrino helicities, $\sigma_x = 1/2\sigma_p $ is the wave packet width in the coordinate space and $N$ is the normalization factor.

Finally, for the probabilities of the neutrino flavour and spin oscillations in a magnetic field we get:
\begin{eqnarray}\label{prob_1}
	P_{\nu_{\alpha}^h \to \nu_{\beta}^{h'}}(L,t) &=& \Big| \sum_{i} U^*_{\beta i} U_{\alpha i} (u_i^{h'})^\dag\nu_i^h(L,t)\Big|^2 \\
	\nonumber
	&=& \frac{1}{N^2} \sum_{i,j} \sum_{s,\sigma=\pm1} U^*_{\beta i} U_{\alpha i} U_{\beta j} U^*_{\alpha j} C_{is}^{hh'} C_{j\sigma}^{hh'} e^{-i\omega_{ij}^{s\sigma}(p_0) t} \exp \Bigg( -\frac{(\phi_i^s)^2 + (\phi_j^{\sigma})^2}{4\sigma_x^2} \Bigg),
\end{eqnarray}
where $\phi_i^s = L-v_i^s(p_0)t$, $\omega_{ij}^{s\sigma}(p_0) = E_i^s(p_0) - E_j^{\sigma}(p_0)$ are the frequencies of neutrino oscillations that are given by
\begin{equation}\label{freq}
	\omega_{ij}^{s\sigma}(p_0)\approx \frac{\Delta m^2_{ij}}{2p_0} - (\mu_i s - \mu_j \sigma)B_{\perp}.
\end{equation}

Note that the oscillations on both the vacuum $\omega_{ij}^{vac}=\Delta m^2_{ij}/2p_0$ and magnetic $\omega_{i}^{B}=\mu_i B_{\perp}$ frequencies present in the oscillations probabilities (\ref{prob_1}).

Since the time $t$ of the neutrino propagation from the  source to the  detector is not an observable quantity, we perform integration over time to obtain the final expression for probabilities of high-energy neutrino oscillations in a magnetic field and get:
\begin{equation}\label{prob_2}
	P_{\nu_{\alpha}^h \to \nu_{\beta}^{h'}}(L) = \sum_{i,j} \sum_{s,\sigma} U^*_{\beta i} U_{\alpha i} U_{\beta j} U^*_{\alpha j} C_{is}^{hh'} C_{j\sigma}^{hh'} \exp\Big( -i2\pi \frac{L}{L_{osc}^{ij s\sigma}}\Big) \exp\Big( -\frac{L^2}{ (L_{coh}^{ij s\sigma})^2} \Big),
\end{equation}
where the corresponding oscillations and coherence lengths are given by

\begin{eqnarray}
	\label{osc_lenght}
	L_{osc}^{ij s\sigma} &=& \frac{\pi}{\omega_{ij}^{s\sigma}}, \\
	\label{coh_lenght}
	L_{coh}^{ij s\sigma} &=& \frac{2\sqrt{2}\sigma_x}{v_i^s-v_j^{\sigma}}.
\end{eqnarray}

Under realistic assumptions that $p_0 \gg m_i \gg \mu_i B$, the approximate expressions for the coherence lengths can be obtained:

\begin{equation}
	L_{coh}^{ijss} \approx  \frac{4\sqrt{2}\sigma_x p^2_0}{\Delta m_{ij}^2}, \;\;\;
	L_{coh}^{ii-+} \approx \frac{2 B_\perp^2}{ B^2 + 3 B_\perp^2}\frac{\sigma_x p^3_0}{\mu_iB_\perp m_i^2}, \;\;\;
	L_{coh}^{ij-+} \approx L_{coh}^{ijss}.
\end{equation}
Note that unlike the coherence lengths $L_{coh}^{ijss}$ for oscillations on vacuum frequencies $\omega_{ij}^{vac} = \Delta m^2_{ij}/4p_0$, the coherence lengths $L_{coh}^{ii-+}$ are proportional to $p^3_0$.

For the case of flavour oscillations Eq.(\ref{prob_2}) can be rewritten as
\begin{eqnarray}\label{prob_B_final}
	\nonumber
	P_{\alpha\beta}(L) &=& P_{\nu_{\alpha}^L \to \nu_{\beta}^{L}}(L) = \frac{1}{2} \sum_{i=1}^3 |U_{\alpha i}|^2 |U_{\beta i}|^2\left[1 + \cos\Big(\frac{2\pi L}{L_i^B}\Big) D_{i}^B (L) \right] \\ 
	&+& 2 \sum_{i>j} \text{Re}(U_{\beta i}^*U_{\alpha i}U_{\beta j}U_{\alpha j}^*) \cos\Big(\frac{2\pi L}{L_{ij}^{vac}}\Big) \cos\Big(\frac{2\pi L}{L_i^B}\Big)\cos\Big(\frac{2\pi L}{L_j^B}\Big) D_{ij}^{vac}(L) \\ \nonumber
	&+& 2 \sum_{i>j} \text{Im}(U_{\beta i}^*U_{\alpha i}U_{\beta j}U_{\alpha j}^*) \sin\Big(\frac{2\pi L}{L_{ij}^{vac}}\Big) \cos\Big(\frac{2\pi L}{L_i^B}\Big)\cos\Big(\frac{2\pi L}{L_j^B}\Big) D_{ij}^{vac}(L),
\end{eqnarray}
where the oscillations lengths are given by
\begin{equation}
	L_{ij}^{vac} = L_{osc}^{ijss} = \frac{\pi p}{\Delta m^2_{ij}}, \;\;\; L_{i}^{B} = L_{osc}^{iis\sigma} = \frac{\pi}{\mu_i B_\perp}
\end{equation}
and
\begin{equation}
	D_{ij}^{vac}(L) = \exp \Big(-\frac{L^2}{(L^{ijss}_{coh})^2} \Big), \;\;\; D_i^B(L) = \exp \Big(-\frac{L^2}{(L_{coh}^{iis\sigma})^2}\Big)
\end{equation}
are the damping terms.
\\
\\
For the probability of spin-flip we get

\begin{equation}\label{spin_flip}
	P_{\nu_{\alpha}^L \to \nu^R}(L) = \sum_\beta P_{\nu_{\alpha}^L \to \nu_{\beta}^{R}}(L) = \sum_{i=1}^3 |U_{\alpha i}|^2 (1 - \cos^2(\pi L/L_i^B) D^B_i(L)).
\end{equation}

The oscillations probabilities (\ref{prob_B_final}) and (\ref{spin_flip}) generalize the expressions from \cite{Kurashvili:2017zab,Lichkunov:2022mjf} and account for exponential damping of neutrino oscillations at large distances due to the wave packets separation.

Due to the unitarity of the neutrino evolution operator, Eq.(\ref{prob_B_final}) and q.(\ref{spin_flip}) satisfy the probability conservation relation

\begin{equation}
	\sum_\beta P_{\alpha\beta}(L) + P_{\nu_{\alpha}^L \to \nu^R}(L) = 1.
\end{equation}

From Eq.(\ref{prob_B_final}) it follows that the oscillations on frequencies that depend on the magnetic field strength $B$ are suppressed by \emph{two different} damping terms $D_{ij}^{vac}(L)$ and $D_{i}^B(L)$. First of them contains the coherence length $L^{ijss}_{coh}$ that coincide with the coherence lengths of neutrino oscillations in vacuum. The second damping term contains coherence lengths $L^{iis\sigma}_{coh}$ that differ from the coherence lengths of neutrino oscillations in vacuum.
\\
\\
Using (\ref{group_vel_expansion}), we obtain the following numerical estimations for the coherence lengths:
\begin{eqnarray}\label{L_coh_num}
	L_{coh}^{ijss} / \sigma_x &\approx& 2\sqrt{2} \left(\frac{p}{1\text{ TeV}}\right)^2  \Big( \frac{\Delta m_{ij}^2}{10^{-5}\text{ eV}^2} \Big)^{-1} 10^{29}, \\
	L_{coh}^{ii-+} / \sigma_x &\sim& \left(\frac{p}{1\text{ TeV}}\right)^3 \left( \frac{m_i}{1\text{ eV}} \right)^{-2} \Big(\frac{B}{1\;\mu\text{G}}\Big)^{-1} \left( \frac{\mu_i}{10^{-11}\mu_B} \right)^{-1} 10^{60}.
\end{eqnarray}

The oscillations lengths are given by
\begin{equation}
	L^{B}_{i} = \frac{\pi}{\mu_i B_{\perp}} =  2.17 \cdot \left( \frac{B}{1\;\mu\text{G}}\right)^{-1} \left( \frac{\mu_i}{10^{-11}\mu_B} \right)^{-1} 10^{3} \; \text{pc},
\end{equation}
\begin{equation}
	L^{vac}_{ij} = \frac{4 \pi p}{\Delta m^2_{ij}} = 5.02 \cdot \left( \frac{\Delta m_{ij}^2 }{10^{-5}\;\text{eV}^2}\right)^{-1} \left(\frac{p}{1\;\text{TeV}} \right) \cdot 10^{-5} \; \text{pc}.
\end{equation}

\begin{figure}[tbp]
	
	\centering 
	\includegraphics[width=.49\textwidth]{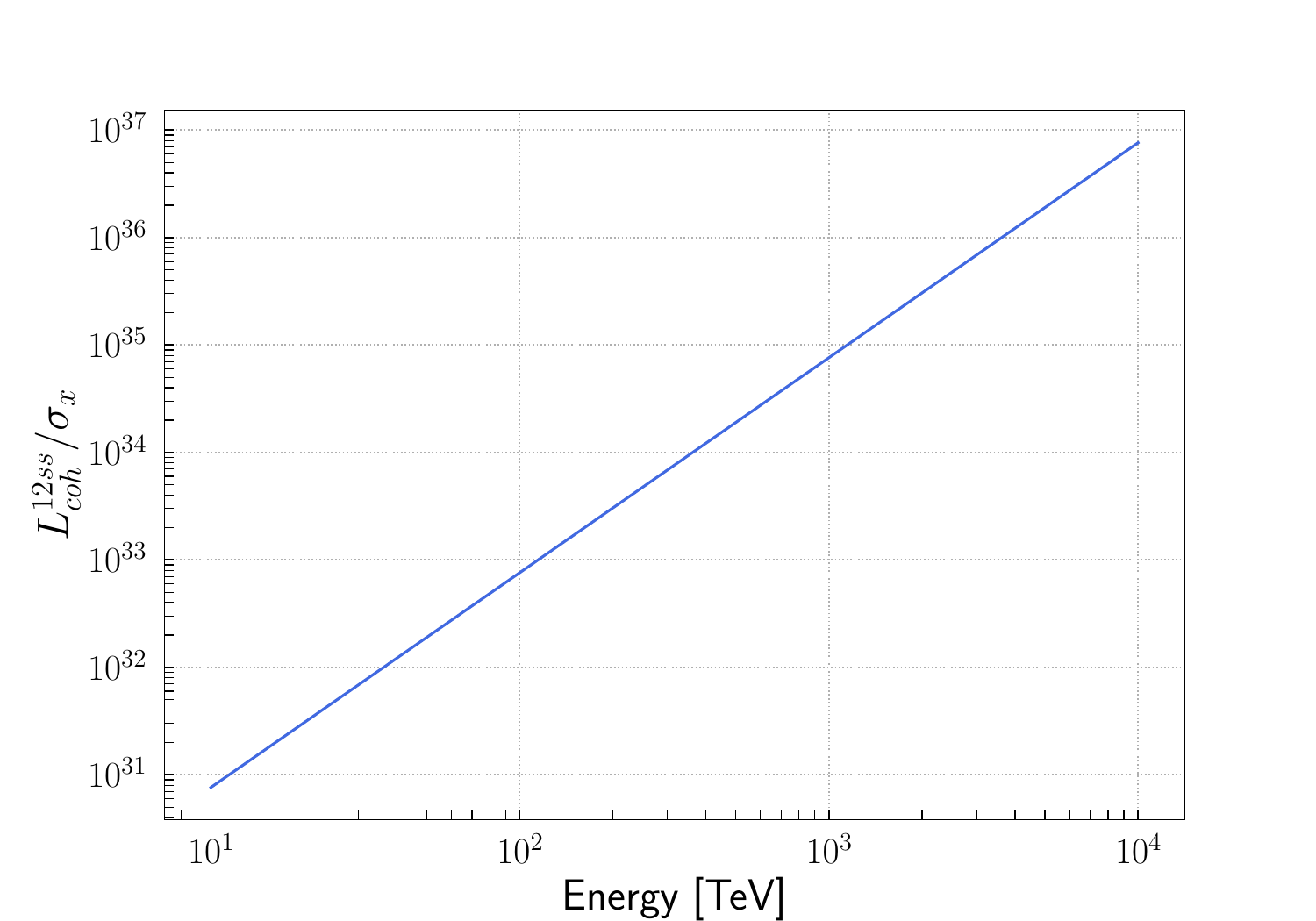}
	\includegraphics[width=.49\textwidth]{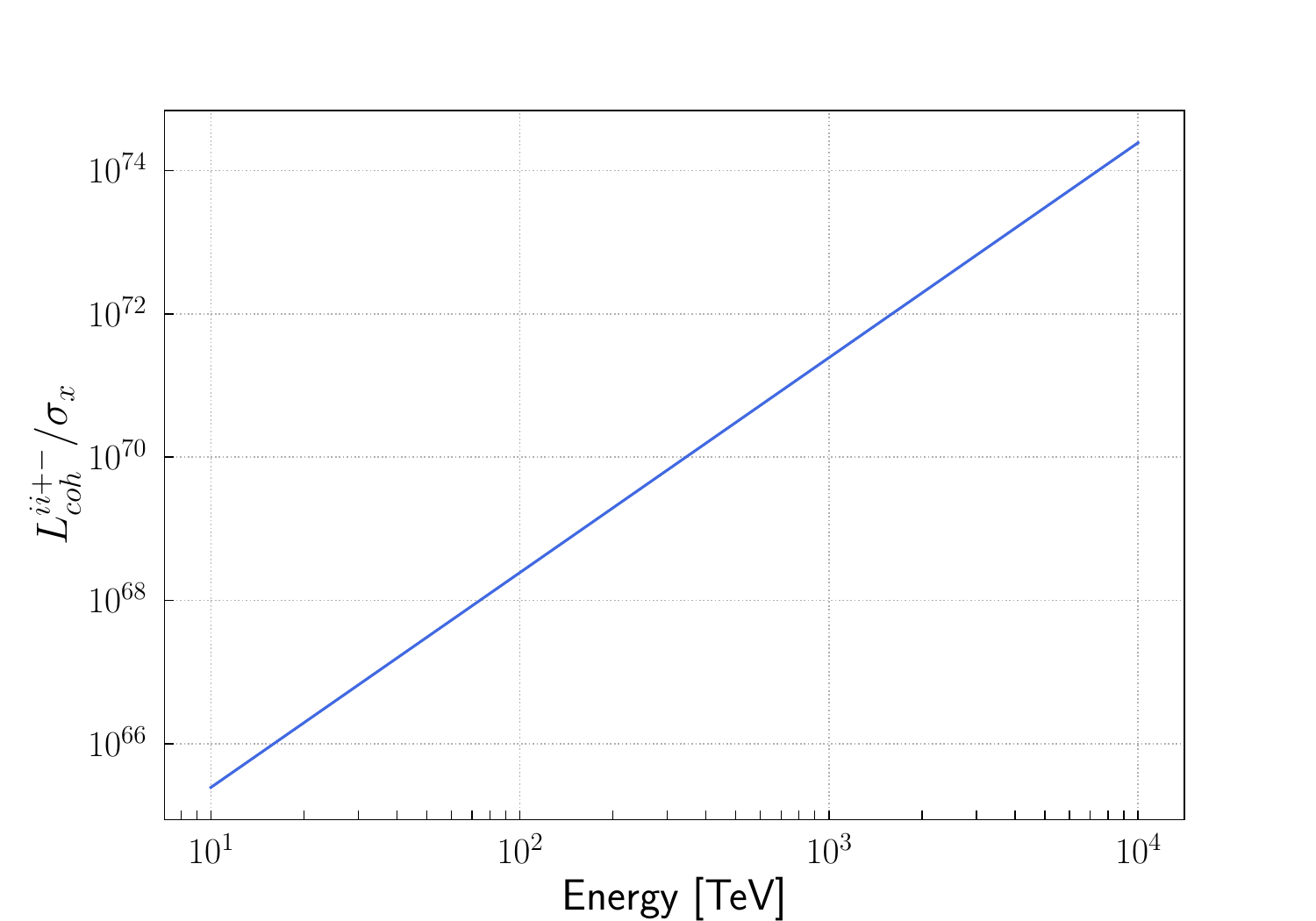}
	\caption{\label{fig:lengths} Dimensionless coherence length $L_{coh}/\sigma_x$. Left: coherence length of oscillations on vacuum frequency $L_{coh}^{12ss}/\sigma_x = \dfrac{4\sqrt{2} p^2_0}{\Delta m_{21}^2}$. Right: coherence length of oscillations on magnetic frequency $L_{coh}^{ii+-}/\sigma_x$ assuming $m_i = 0.1$ eV, $\mu_i = 10^{-12} \mu_B$ and $B = 10^{-6}$ Gauss.}
\end{figure}

The dimensionless coherence lengths $L_{coh}^{ijss} / \sigma_x$ and $L_{coh}^{ii+-} / \sigma_x$ are shown in Figure \ref{fig:lengths}. For the coherence length of oscillations on magnetic frequency $L_{coh}^{ii+-}$ we assume $m_i = 0.1$ eV, $\mu_i = 10^{-12} \mu_B$ and $B = 10^{-6}$ Gauss.

There are various estimations for neutrino wave packet width $\sigma_x$ in literature. The experimental limits on $\sigma_x$ for neutrinos from the terrestrial experiments are obtained in \cite{DayaBay:2016ouy,deGouvea:2021uvg,deGouvea:2024syg,Smolsky:2024uby}. Theoretical calculation of the neutrino wave packet width can be also found in \cite{Akhmedov:2022bjs,Jones:2022hme}.

For the astrophysical high-energy neutrinos, it is shown in \cite{Farzan:2008eg} that for neutrinos produced by the free $\pi^\pm$ or $\mu^\pm$ decays $\sigma_x$ are given by
\begin{eqnarray}\label{sigma_x}
	\sigma_x &\sim& 10^{-3} \left( \frac{10\text{ TeV}}{E_\nu} \right)\text{ cm} \;\;\;\;\;(\text{for $\pi^{\pm}$ decay}), \\
	\sigma_x &\sim& 10^{-1} \left( \frac{10\text{ TeV}}{E_\nu} \right)\text{ cm} \;\;\;\;\;(\text{for $\mu^{\pm}$ decay}).
\end{eqnarray}

The appearance of the magnetic field in the neutrino source may decrease the wave packet width by orders of magnitude \cite{Farzan:2008eg}. In \cite{Akhmedov:2017mcc}, the authors show that for neutrinos produces in the supernova explosions $\sigma_x \sim 10^{-12}$ cm. Assuming that high-energy neutrinos are produced by the decay of free $\pi^\pm$ the coherence length is given by (\ref{sigma_x}), and for $100$ TeV neutrinos we obtain $L_{coh}^{13ss} \sim 10^9$ pc and $L_{coh}^{12ss} \sim 10^{11}$ pc.

Currently, the most distant source of neutrinos is the blazar TXS 0506+056 \cite{IceCube:2018-blazar} located at approximately 1.7 Gpc. From our estimations it follows that oscillations on the vacuum frequencies can disappear due to the wave packet separation for astrophysical high-energy neutrinos coming from the most distant sources in the Universe. Assuming $B = 1\mu$G, $\mu_i = 10^{-12}\mu_B$ and $m_i = 0.45$ eV, for the coherence lengths of oscillations on the magnetic frequencies $\omega_i^B = \mu_i B_{\perp}$ we obtain $L_{coh}^{ii-+} \sim 10^{46}$ pc, that by orders of magnitudes exceeds the scale of observable universe which is approximately 30 Gpc.

Thus, for baselines $L \gg L^{vac}_{coh} = \max{\big(L_{coh}^{12ss}, L_{coh}^{13ss}\big)}$ \emph{partial damping} of neutrino oscillations occurs, and for the damping terms we have $D_{ij}^{vac} \approx 0$ and $ D_i^B \approx 1$. We obtain the following expressions for the probabilities of flavour and spin oscillations at distances exceeding the coherence lengths of neutrino oscillations in vacuum $L^{vac}_{coh} = \max{\big(L_{coh}^{12ss}, L_{coh}^{13ss}\big)}$:
\begin{eqnarray}
	P_{\alpha \beta}(L)\Big|_{L \gg L_{coh}^{vac}} = \sum_{i=1}^3 |U_{\alpha i}|^2|U_{\beta i}|^2 \cos^2 (\pi L/L_i^B), \\
	P_{\nu_{\alpha}^L \to \nu^R}(L)\Big|_{L \gg L_{coh}^{vac}} = \sum_{i=1}^{3} |U_{\alpha i}|^2 \sin^2(\pi L/L_i^B).
\end{eqnarray}

\begin{figure}[tbp]
	\centering 
	\includegraphics[width=.49\textwidth]{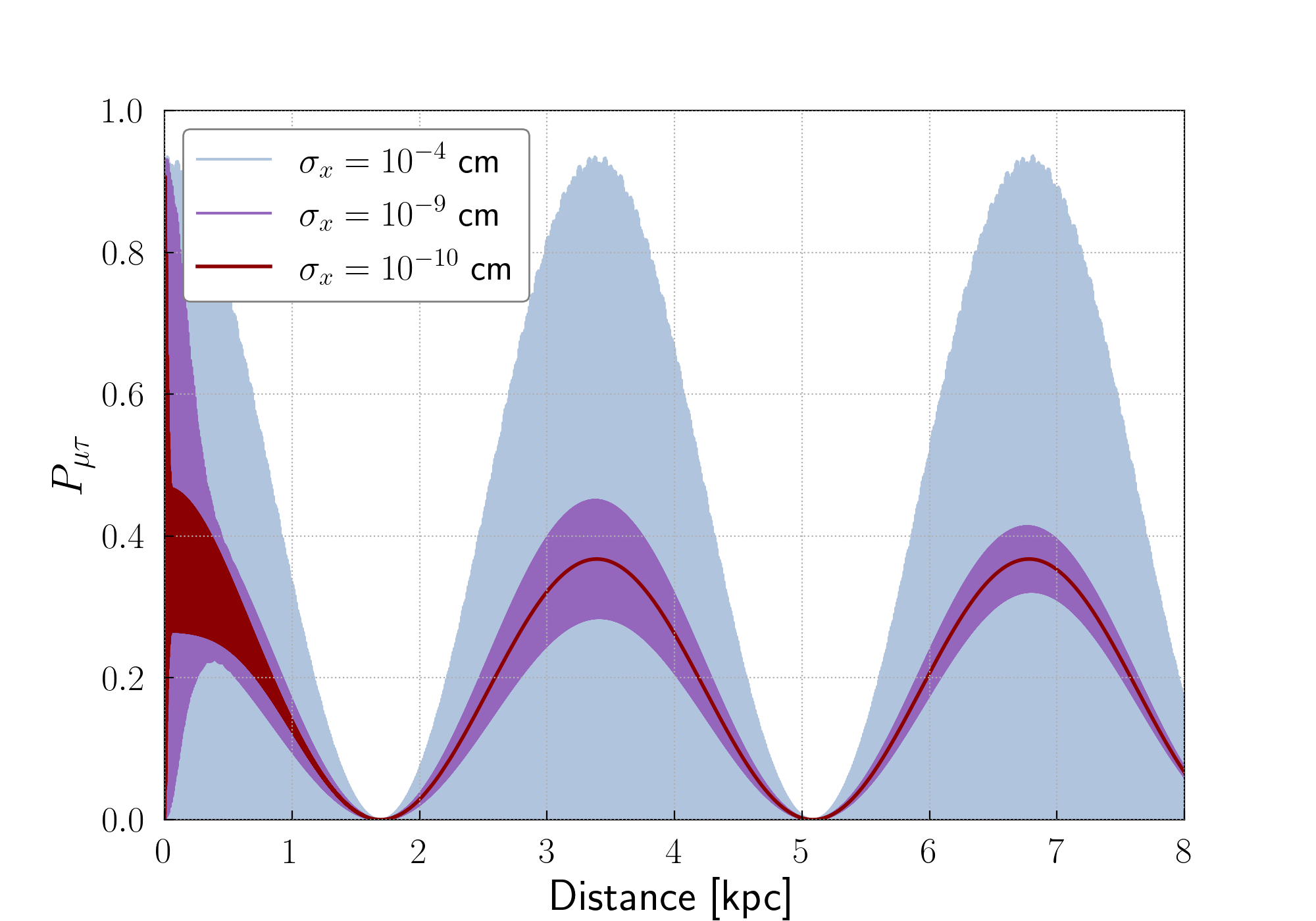}
	\includegraphics[width=.49\textwidth]{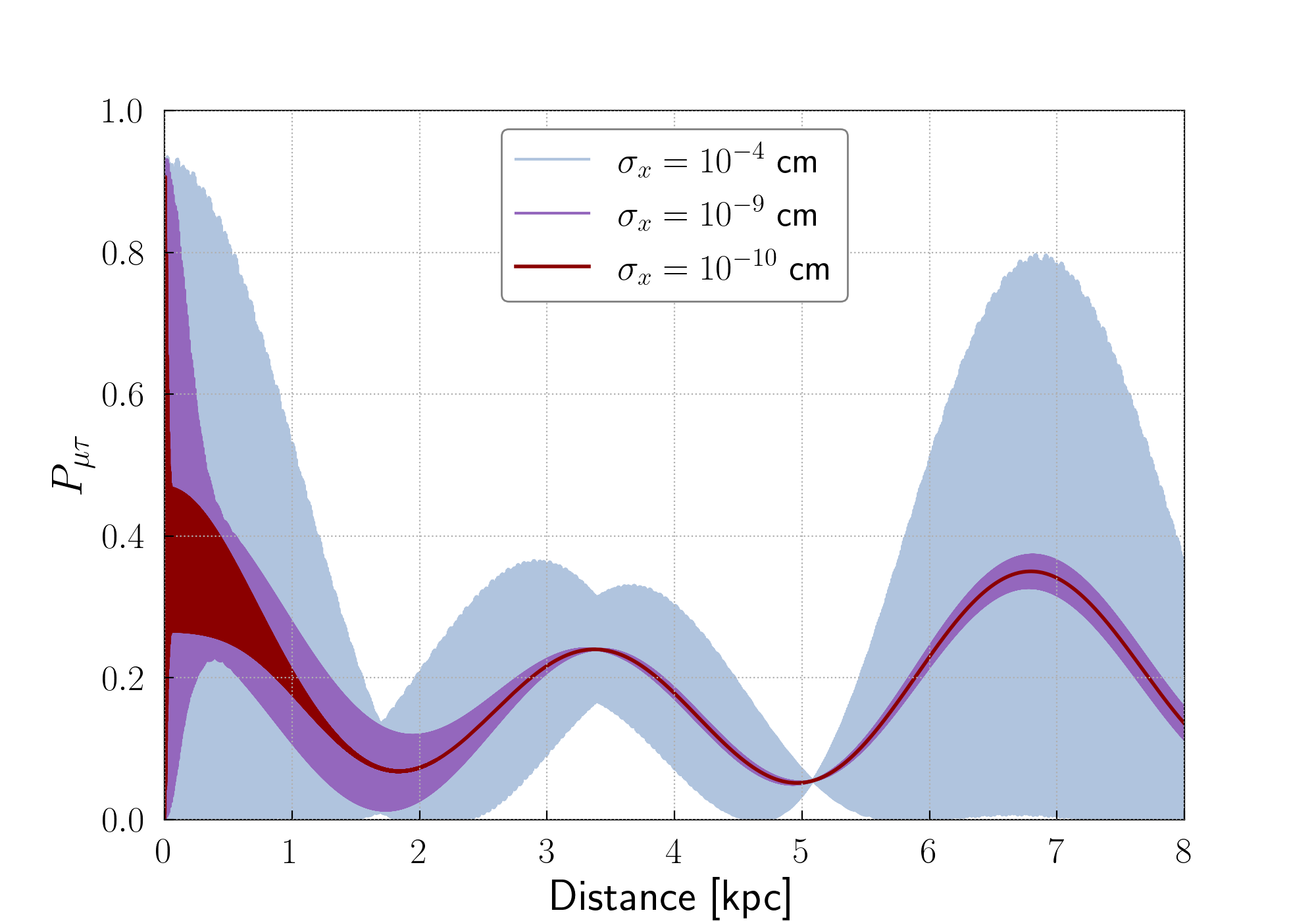}
	\caption{\label{fig:prob} The probability of neutrino flavour oscillations $\nu_\mu \to \nu_\tau$ in a magnetic field. Left: $(\mu_1,\mu_2,\mu_3) = (1,1,1)\times 6.4\cdot10^{-12}\mu_B$. Right: $(\mu_1,\mu_2,\mu_3) = (1/3,1/2,1)\times6.4\cdot10^{-12}\mu_B$.}
\end{figure}

The phenomenon of partial decoherence is illustrated in Figure \ref{fig:prob}, where we plot the neutrino flavour oscillations probability $P_{\mu\tau}$ in the Galactic magnetic field $B \approx 1$ $\mu$G for the average neutrino momentum $p_0 = 100$ TeV for different values of the neutrino magnetic moments and neutrino wave packet width $\sigma_x$. The filled regions in Figure \ref{fig:prob} correspond to fast oscillations with the characteristic lengths given by $L^{vac}_{12} = 6.8\cdot 10^{-4}$ pc and $L^{vac}_{13} = 5.0\cdot 10^{-6}$ pc, which are modulated by oscillations on the magnetic lengths $L^{B}_{i} \sim 10^{3}$ pc. For $\sigma_x = 10^{-4}$ cm decoherence does not occ. In this case the oscillations lengths are given by $L^{vac}_{12} = 6.8\cdot 10^{-4}$ pc, $L^{vac}_{13} = 5.0\cdot 10^{-6}$ pc and $L^{B}_{i} \sim 10^{3}$ pc. For $\sigma_x = 10^{-4}$ cm decoherence does not occur on the considered 8 kiloparsec scale. Oscillations characterized by the oscillations length $L_{13}^{vac}$ disappear after few hundred parsecs for $\sigma_x = 10^{-9}$ cm, while oscillations on the length $L_{12}^{vac}$ only slowly decay. Finally, for $\sigma_x = 10^{-10}$ cm oscillations on vacuum lengths completely disappear after $\approx 1$ kpc, while oscillations on the magnetic lengths $L_i^B$ are not suppressed.

Above we only considered neutrino oscillations in the Galactic magnetic field. For the case of extragalactic neutrino sources the studied effects may be more pronounced due to larger baseline $L$. The considered effects also may be important for studying neutrino oscillation inside astrophysical objects.

\subsection{Neutrino oscillations in a non-uniform magnetic field}
In previous section we considered the neutrino wave packet evolution in a uniform magnetic field. In this case it is possible to obtain analytical expressions for the neutrino oscillations probabilities. However, real astrophysical magnetic fields have the complicated spatial structure. In this section we consider the evolution of the neutrino wave packet in a non-uniform magnetic field.

The neutrino wave function evolution in a magnetic field is described by the following Dirac equation:
\begin{equation}\label{Dirac_eq_WP}
	(i\gamma^{\mu}\partial_{\mu} - m_i)\nu_i(x,t) + \mu_{i}\bm{\Sigma}\bm{B}(x) \nu_i(x,t) = 0.
\end{equation}
As it is shown in Section \ref{sec:3.1}, for the case of a uniform magnetic field the partial differential equation (\ref{Dirac_eq_WP}) can be transformed into an ordinary differential equation (\ref{uniform_B_p_space}) by the transition to the momentum space. However, for the case of a non-uniform magnetic field in the momentum space we obtain the integro-differential equation
\begin{equation}\label{wp_nonuniform_B}
	i\gamma^{0}\partial_{t}\nu_i(p,t) = (\bm{\gamma}\bm{p} + m_i)\nu_i(p,t) - \mu_{i} \int \bm{\Sigma} \bm{B}(p-q) \nu_i(q,t) dq.
\end{equation}

The Eq. (\ref{wp_nonuniform_B}) can be simplified under the assumption that the spacial scale on which the magnetic field varies (the so-called magnetic field \textit{coherence length}\footnote{The magnetic field coherence length not to be confused with  the neutrino oscillations coherence lengths that characterise the damping of the oscillations} $\lambda$) is significantly larger that the wave packet width $\sigma_x$.

As it is given by Eq.(\ref{wave_fun_x_space}), the neutrino wave function has the following form
\begin{equation}\label{WP_avg_x}
	\nu_i^h(x,t) = \frac 1 N \sum_{s,h'} C_{is}^{hh'} e^{-iE_i^s(p_0)t + ip_0 x} \exp\Big({-\frac{(x - \langle x_i^s(t) \rangle)^2}{4\sigma_x^2}}\Big) u_i^{h'},
\end{equation}
where $\langle x^s_i(t) \rangle$ are the average coordinates of the corresponding wave packets for different massive neutrino spin states.
It is also shown in Section \ref{sec:3.1}, that for the case of high-energy neutrinos and realistic values of the interstellar magnetic fields, the loss of the coherence of different neutrino spin states does not occur even on the cosmological scales. Therefore, the following condition
\begin{equation}
	| \langle x^s_i(t) \rangle - \langle x^\sigma_i(t) \rangle | \ll \sigma_x
\end{equation}
is satisfied even for the ultra-long baselines. Thus, it can be assumed that $\langle x^s_i(t) \rangle = \langle x^\sigma_i(t)\rangle = \langle x_i(t) \rangle$ in Eq.(\ref{WP_avg_x}) for all neutrino spin numbers $s,\sigma$.

Next, we decompose the strength of the magnetic field near the average coordinate $\langle x_i(t) \rangle$:
\begin{equation}
	\bm{B}(x) = \bm{B}(\langle x_i(t) \rangle) + (x - \langle x_i(t) \rangle)\cdot \frac{\partial \bm{B}(x)}{\partial x} \Bigg\rvert_{\langle x_i(t) \rangle} + ...
\end{equation}
If the magnetic field remains approximately constant on the wave packet width, the term containing derivative can be neglected. Consider neutrino oscillations in the interstellar magnetic fields that have the coherence length $\lambda \gtrsim$ 10 parsec, which is many orders of magnitude larger that the neutrino wave packet size. Under a realistic assumption that the magnetic field does not vary significantly on the scale of the neutrino wave packet, i.e. $\lambda \gg \sigma_x$, we perform the Fourier transform of (\ref{Dirac_eq_WP}) and arrive to the following equation:
\begin{equation}\label{eq_nonuniform_1}
	i\gamma^{0}\partial_{t}\nu_i(p,t) = (\gamma_3 p + m_i)\nu_i(p,t) - \mu_i  \bm{\Sigma} \bm{B} (\langle x_i(t) \rangle) \nu_i(p,t).
\end{equation}

Eq.(\ref{eq_nonuniform_1}) can be further simplified if we suppose that the magnetic field $\bm{B}$ remains approximately constant on the scale of the distance between the neutrino wave packets. For the ultrarelativistic neutrinos the said distance can be estimated as
\begin{equation}
	\Delta x_{ij} = \frac L c (v_i - v_j) = \frac{\Delta m_{ij}^2}{2p^2} \frac L c,
\end{equation}
where $L$ is the distance from source. For 100 TeV neutrinos propagating for 10 kiloparsec we get: $\Delta x_{12} \sim 10^{-11}$ cm and $\Delta x_{13} \sim 10^{-9}$ cm, while for 10 MeV supernova neutrinos $\Delta x_{12} \sim 1$ km and $\Delta x_{13} \sim 10$ m. This values are significantly lower than the coherence length of the interstellar magnetic field, and consequently we can assume $\bm{B} (\langle x_i(t) \rangle) \approx \bm{B} (t)$.

Thus, we arrive to the following evolution equation for the neutrino wave packet in a magnetic field:
\begin{equation}\label{WP_eq_approximate}
	i \gamma^0 \partial_t \nu_i(p,t) = (\gamma_3 p + m_i) \nu_i(p,t) - \mu_i \bm{\Sigma} \bm{B} (t) \nu_i(p,t) = 0.
\end{equation}

In the next section, Eq.(\ref{WP_eq_approximate}) is solved numerically to describe the neutrino flux evolution in the Galactic and extragalactic magnetic field.

\section{Flavour composition of astrophysical neutrinos}
\label{sec:iv}

\subsection{Flavour composition of neutrinos from point-like sources}
The differential flux of the flavour neutrino $\nu_\beta$ ($\beta = e,\mu,\tau$) observed by a terrestrial neutrino telescope is given by
\begin{equation}
	\frac{d \Phi_{\beta}(L, E)}{dE} = \frac{1}{4\pi L^2} \sum_{\alpha = e,\mu, \tau} \frac{d \Phi^0(E)}{dE} r^0_\alpha(E) P_{\alpha \beta}(L,E),
\end{equation}
where $d\Phi^0/dE$ is the all-flavour differential flux at the source, $r^0_\alpha$ are the neutrino flavour ratios at the source, $L$ is the distance from the neutrino source to the detector and $P_{\alpha\beta}(L,E)$ are the oscillations probabilities given by (\ref{prob_B_final}). We also consider the integrated neutrino fluxes
\begin{equation}\label{flux_integral}
	\Phi_{\beta}(L) = \int_{E_{min}}^{E_{max}} \frac{d \Phi_{\beta}(L,E)}{dE} dE.
\end{equation}
For the current high-energy neutrino telescopes IceCube, Baikal-GVD and KM3NeT integration is performed within the energy range between $E_{min} \approx 100$ TeV and $E_{max} \approx 10$ PeV. However, it is possible that in future experiments neutrinos with energies $10^{19}$ eV and above will be detected \cite{Ackermann:2022rqc}.

In our calculations we use the single power-law model of the neutrino flux
\begin{equation}
\dfrac{d\Phi^0(E)}{dE}  \sim E^{-\gamma},
\end{equation}
where the spectral index $\gamma \approx 2.5$. This model is currently consistent with the flux observed by IceCube and Baikal-GVD.

For the case of energy-independent initial flavour ratios $r^0_\beta$, that are used in the IceCube data analysis, for the flavour ratios we have

\begin{equation}\label{flavour_ratios}
r_\beta(L) = \frac{\Phi_\beta}{\sum_\beta \Phi_\beta} = \frac{\sum_\alpha r^0_\alpha \langle P_{\alpha\beta}(L,E) \rangle_E }{\sum_\alpha r^0_\alpha (1-P_{\nu_{\alpha}^L \to \nu^R}(L)) },
\end{equation}

where the oscillations probabilities averaged over neutrino energy are given by

\begin{eqnarray}\label{prob_int}
\nonumber
\langle P_{\alpha\beta}(L,E) \rangle_E &=& \frac{1-\gamma}{E_{max}^{1-\gamma} - E_{min}^{1-\gamma}} \int_{E_{min}}^{E_{max}} E^{-\gamma} P_{\alpha\beta}(L,E) dE = \sum_{i=1}^3 |U_{\alpha i}|^2 |U_{\beta i}|^2 \cos^2\Big(\frac{\pi L}{L_i^B}\Big) D_{i}^B (L) \\ 
&+& 2 \sum_{i>j} \text{Re}(U_{\beta i}^* U_{\alpha i}U_{\beta j}U_{\alpha j}^*) \cos\Big(\frac{2\pi L}{L_i^B}\Big)\cos\Big(\frac{2\pi L}{L_j^B}\Big) \tilde{C}_{ij}(L) \\ \nonumber
&+& 2 \sum_{i>j} \text{Im}(U_{\beta i}^* U_{\alpha i}U_{\beta j}U_{\alpha j}^*) \cos\Big(\frac{2\pi L}{L_i^B}\Big)\cos\Big(\frac{2\pi L}{L_j^B}\Big) \tilde{S}_{ij}(L),
\end{eqnarray}
where
\begin{eqnarray}\label{c}
\tilde{C}_{ij}(L) &=& \frac{1-\gamma}{E_{max}^{1-\gamma} - E_{min}^{1-\gamma}} \int_{E_{min}}^{E_{max}} E^{-\gamma} \cos\Big( \frac{\Delta m^2_{ij} L}{2 E} \Big) D_{ij}^{vac}(L,E) dE,\\
\label{s}
\tilde{S}_{ij}(L) &=& \frac{1-\gamma}{E_{max}^{1-\gamma} - E_{min}^{1-\gamma}} \int_{E_{min}}^{E_{max}} E^{-\gamma} \sin\Big( \frac{\Delta m^2_{ij} L}{2 E} \Big) D_{ij}^{vac}(L,E) dE
\end{eqnarray}
are the terms characterizing the contribution from the oscillations on vacuum frequencies $\omega_{ij}^{vac} = \Delta m^2_{ij}/2E$ to the neutrino flavour composition.

In general, integrals (\ref{c}) and (\ref{s}) are computed numerically. Analytical estimations are possible in the two limiting cases: $D_{ij}^{vac} \approx 1$, and $D_{ij}^{vac} \approx 0$ (in this case $\tilde{C} = \tilde{S} =0$). Assuming $\gamma-1 >0$ and $D_{ij}^{vac} \approx 1$, one can get
\begin{eqnarray}\label{C_approx}
	\tilde{C}_{ij}(L) &=& \frac{1-\gamma}{E_{max}^{1-\gamma} - E_{min}^{1-\gamma}} \Big(\frac{\Delta m^2_{ij} L}{2}\Big)^{1-\gamma} \text{Im} (i^\gamma\Gamma(\gamma-1, -i\varphi_{ij}(E))\big\rvert_{E_{min}}^{E_{max}}), \\
	\tilde{S}_{ij}(L) &=& -\frac{1-\gamma}{E_{max}^{1-\gamma} - E_{min}^{1-\gamma}} \Big(\frac{\Delta m^2_{ij} L}{2}\Big)^{1-\gamma} \text{Re} (i^\gamma\Gamma(\gamma-1, -i\varphi_{ij}(E))\big\rvert_{E_{min}}^{E_{max}}),
\end{eqnarray}
where $\Gamma(s,z)$ is the incomplete gamma function and $\varphi_{ij}(E) = \Delta m^2_{ij} L/2E$ are the vacuum oscillations phases.

The first term in (\ref{prob_int}) corresponds to the oscillations on the magnetic frequencies $\omega_i = \mu_i B_\perp$, while two last term describe the effect of interplay of oscillations on the vacuum and the magnetic frequencies discussed in \cite{Kurashvili:2017zab,Popov:2019nkr,Lichkunov:2022mjf}. To estimate the relative contribution to the measured neutrino flavour composition from the oscillations the on vacuum frequencies $\Delta m^2_{ij}/2E$, we rewrite (\ref{C_approx}) as
\begin{equation}\label{I_ij}
\tilde{C}_{ij}(E_{min},{E_{max}},L,\gamma) = \frac{\gamma -1}{1 - (\frac{E_{max}}{E_{min}})^{1-\gamma} } \varphi^{1-\gamma}(E_{min}) \text{Im} (i^\gamma\Gamma(\gamma-1, -i\varphi_{ij}(E))\big\rvert_{E_{min}}^{E_{max}}).
\end{equation}
It can be shown than at large distances $L$ integral $\tilde{C}_{ij}$ behaves as $\sim \varphi^{-1}(E_{min})$. Thus, for large oscillations phases $\varphi(E_{min}) \gg 1$ we can safely neglect two last terms in (\ref{prob_int}). In this case the terms containing the damping factors $D^{vac}_{ij}$ disappear, but the terms containing $D^{B}_i$ remain. The flavour composition then does not depend on the spectral index $\gamma$ and is simply described by

\begin{equation}\label{ratios_final}
	r_\beta (L) = \frac{\sum_\alpha r^0_\alpha \sum_{i=1}^3 |U_{\alpha i}|^2|U_{\beta i}|^2 \cos^2 (\pi L/L_i^B) D^{B}_i(L) }{\sum_\alpha r^0_\alpha \sum_{i=1}^3 |U_{\alpha i}|^2 \cos^2 (\pi L/L_i^B) D^{B}_i(L)} .
\end{equation}

\begin{figure}[tbp]
	\centering 
	\includegraphics[width=.49\textwidth]{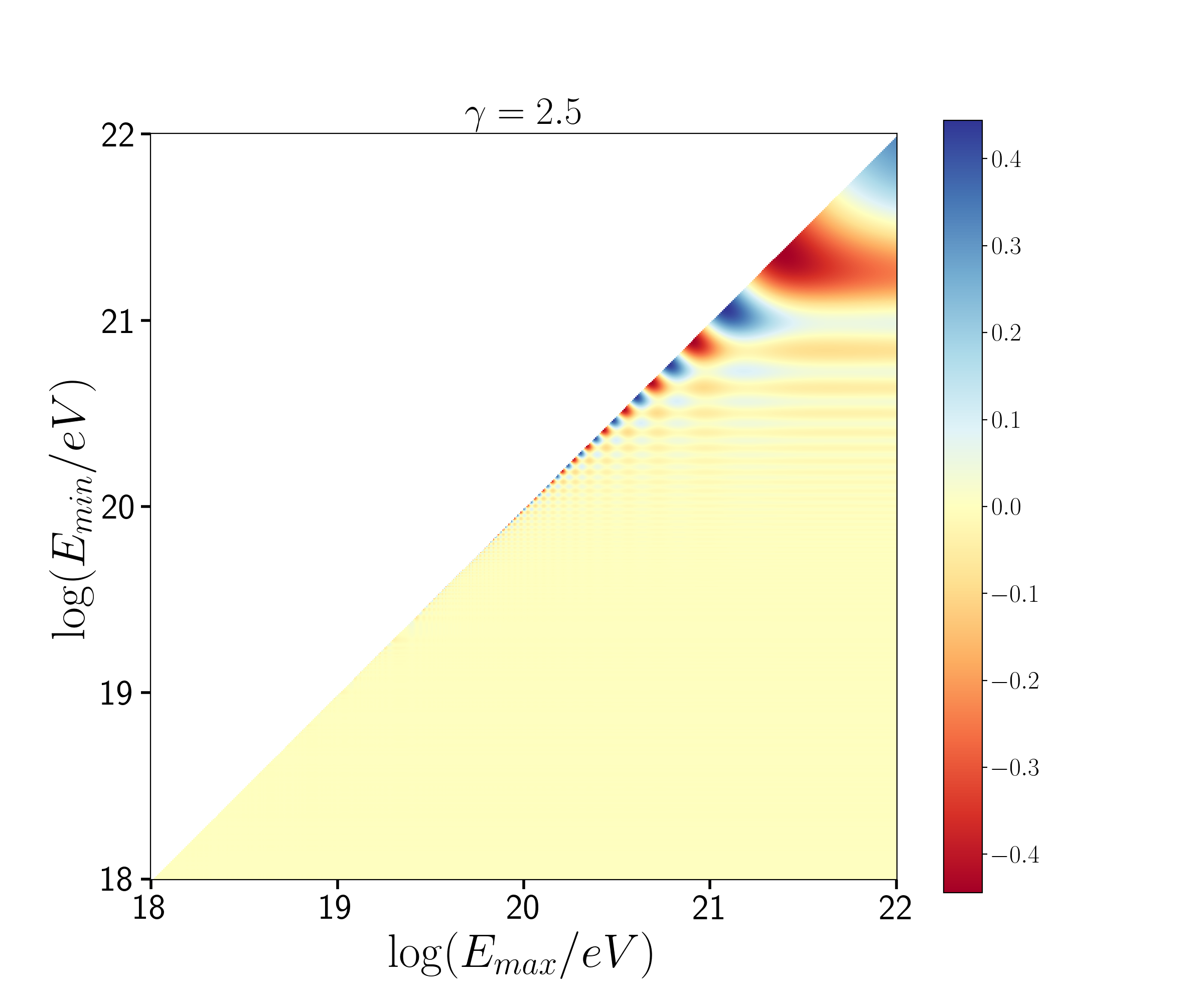}
	\includegraphics[width=.49\textwidth]{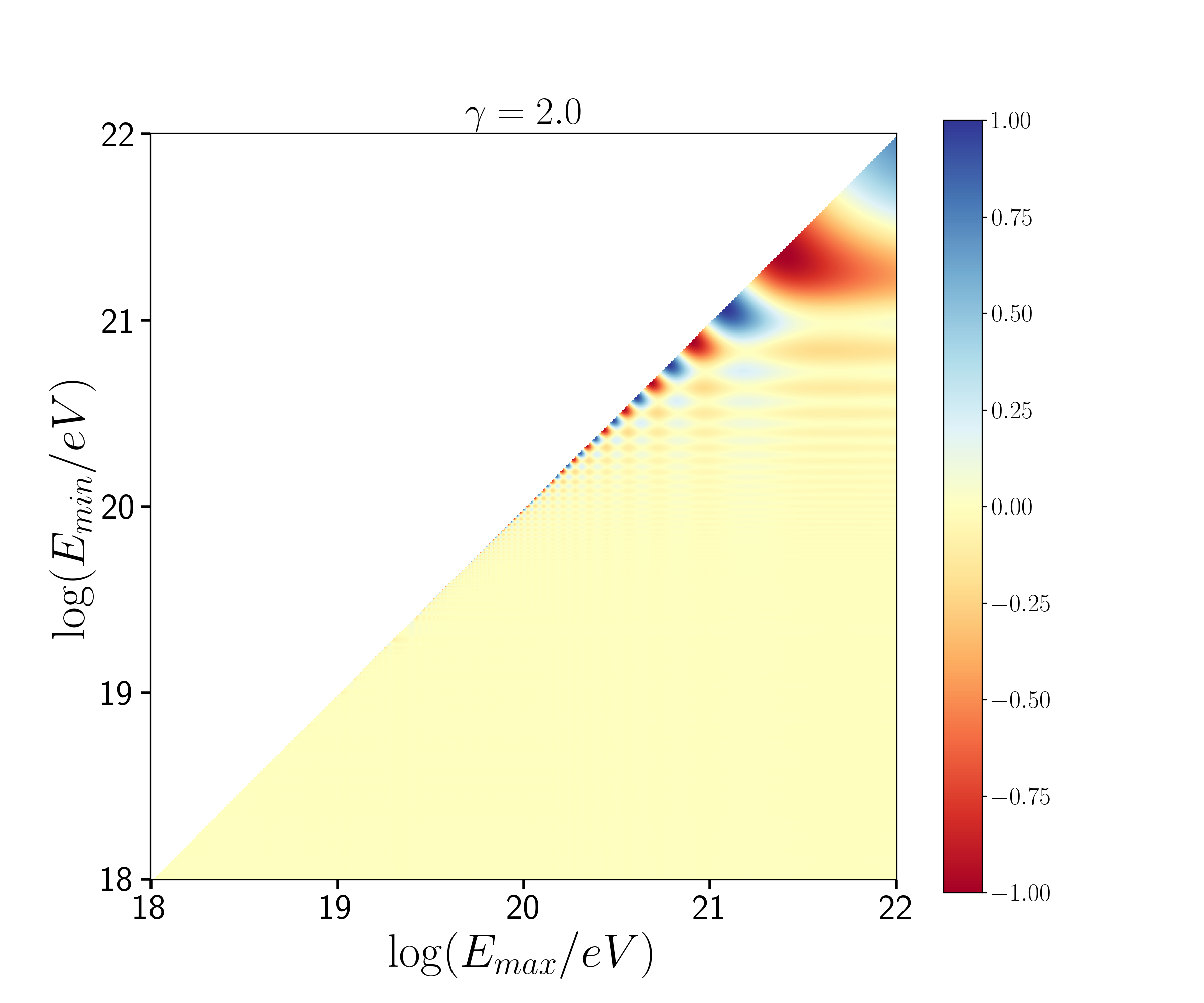}
	\caption{\label{fig:integral} The integral $\tilde{C}_{12}$ assuming distance to the neutrino source $L=8$ kpc for different values of minimal and maximal neutrino energy $E_{min}$ and $E_{max}$. Left: spectral index $\gamma = 2.5$. Right: spectral index $\gamma = 2.0$.}
\end{figure}

Figure \ref{fig:integral} shows $\tilde{C}_{12}$, that characterizes relative contribution to the neutrino flavour composition from oscillations on vacuum frequency $\Delta m^2_{21}/2E$, as the function of minimal and maximal neutrino energy $E_{min}$ and $E_{max}$, assuming $L = 8$ kpc, for spectral indices $\gamma=2.5$ and $\gamma = 2$. This contribution turns out to be non-negligible only for ultra-high neutrino energies ($10^{20}$ eV and higher). This is in accordance with Eq. (\ref{I_ij}), since for neutrino energies $\sim 10^{20}$ eV the oscillations phase $\varphi_{12} = \Delta m^2_{21}L/2E \sim 0.1$ for $L = 8$ kiloparsec. As follows from Eq. (\ref{I_ij}), the smaller the spectral index $\gamma$, the bigger $\tilde{C}_{12}$. $\tilde{S}_{12}$ that characterizes the amplitudes of the $CP$- and $T$-odd terms in the oscillations probability (\ref{prob_int}) can be analyzed in the same way.

Thus, the main conclusion of this section is that averaging over observed neutrino energy results in damping of contributions from oscillations on the vacuum frequencies $\omega_{ij}^{vac} =\Delta m^2_{ij}/2E$ to the neutrino flavour composition, but does not suppress oscillations on the magnetic frequencies $\omega_{i}^{B} =\mu_i B_\perp$. On a phenomenological level, this suppression is similar to the suppression caused by the wave packets separation, but it occurs due to completely different physical mechanisms. The absence of suppression of the oscillations on the magnetic frequencies is due to independence of these frequencies on the neutrino energy. The contribution from the oscillations on the vacuum frequencies to the neutrino flavour compositions becomes significant only for neutrino energies of $10^{20}$ eV and higher. In this case damping of the terms in (\ref{prob_int}) that describe the interplay between the oscillations on the vacuum and magnetic frequencies are only subjected to the damping caused by the wave packets separation, and the flavour composition depends on the neutrino wave packet width.

The results of this section are analytical in nature, and are valid for the case of uniform magnetic field or slowly varying magnetic field, when the adiabatic approximation is justified. In the next two sections we numerically study the neutrino wave packets evolution in a non-uniform magnetic field by adopting a realistic model of the Galactic magnetic field. Following the conclusions of this section, we separately study the cases of high-energy and ultra high-energy neutrinos.

\subsection{Flavour composition of high-energy neutrinos from the Galactic centre}\label{sec:4.2}

In this section we examine possible effects of nonzero neutrino magnetic moments on flavour composition of the high-energy neutrino fluxes (100 TeV to 10 PeV) coming from specific galactic sources. Although they have not yet been observed, there are a number of proposed high-energy neutrino sources within our Galaxy, such as supermassive black hole Sagittarius A$^*$ located at the Galactic center, supernova remnant RX J1713.7-3946, star forming region Cygnus and others.
Note that since radii of these objects are approximately in the range $\sim 10\div100$ parsec and are much smaller than the oscillations length in the galactic magnetic field, we can consider them as point-like sources.
For studying effects of the neutrino interaction with a magnetic field, we are particularly interested in neutrinos coming from the supermassive black hole Sgr A$^*$ since prior to detection they propagate a long distance in the Galactic magnetic field, namely $L = 8178 \pm 13_{stat} \pm 22_{sys}$ \cite{GC_distance}.

To describe the magnetic field of the Galactic plane, we use the model from \cite{Jansson:2012pc}. Here we consider the neutrino interaction with the regular component of the magnetic field. The interaction with a stochastic magnetic field for the solar neutrinos (see \cite{Semikoz:1998ef}) and for neutrinos propagating in interstellar magnetic field (see  \cite{Kurashvili:2020nwb}) leads to the additional damping of neutrino oscillations. However, this effect arises due to the terms proportional to $\mu_i^2$ and are small for neutrino interacting with the Galactic magnetic field.

The modeling of the high-energy neutrino flavour composition involves a sequence of the following steps:
\begin{enumerate}
	
	\item[1)] the solution Eq.(\ref{WP_eq_approximate}) numerically for different values of neutrino magnetic moments $\mu_i$ and different values of the magnetic field model parameters drawn from $3\sigma$ intervals,
	
	\item[2)] the calculation of the probabilities of neutrino flavour oscillations using particular values of neutrino mixing parameters,
	
	\item[3)] the computation of the flavour ratios using Eq.(\ref{flavour_ratios}) for a particular initial flavour ratios $r^0$.
\end{enumerate}

\begin{figure}[tbp]
	\centering 
	\includegraphics[width=.49\textwidth]{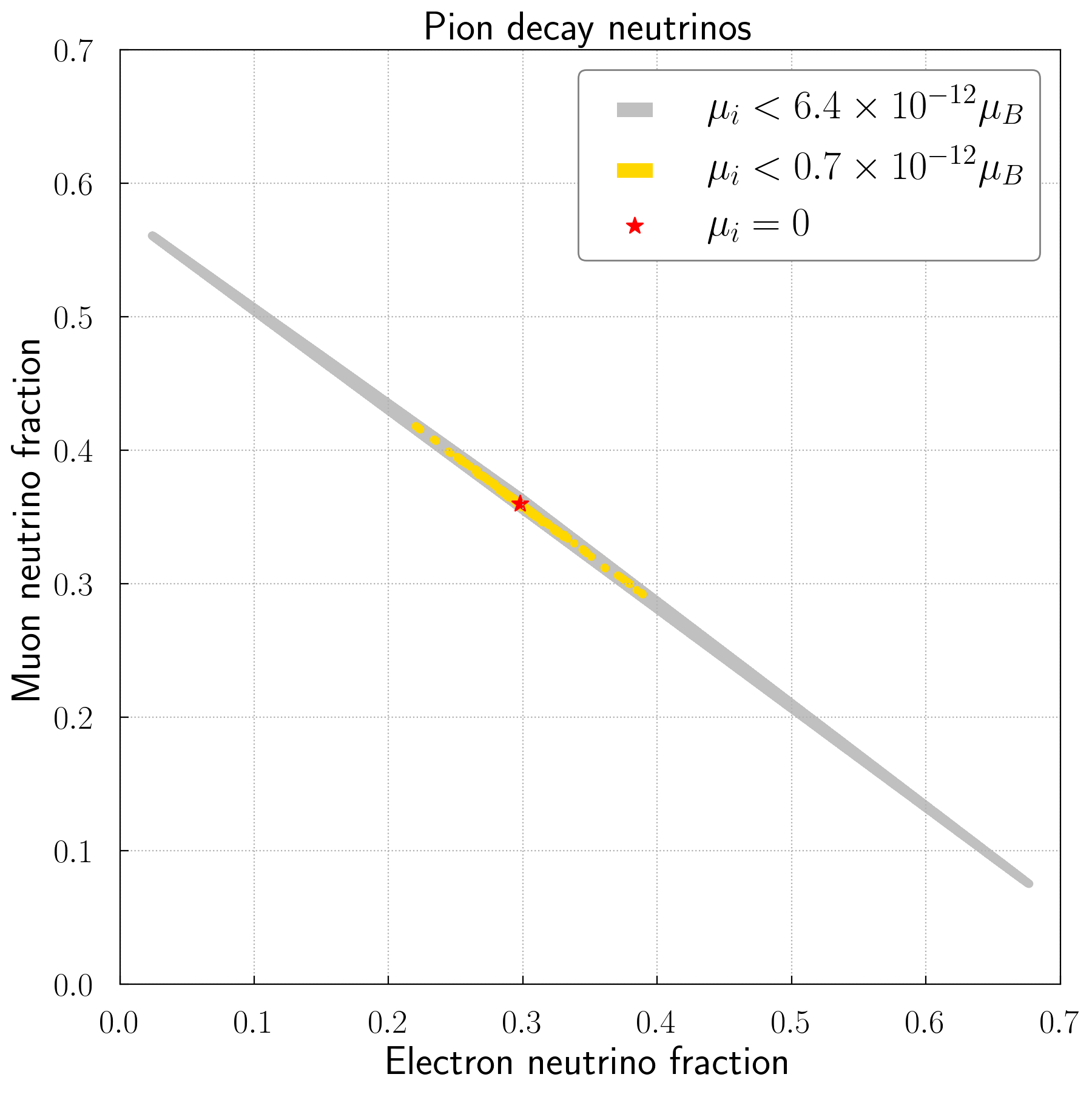}
	\includegraphics[width=.49\textwidth]{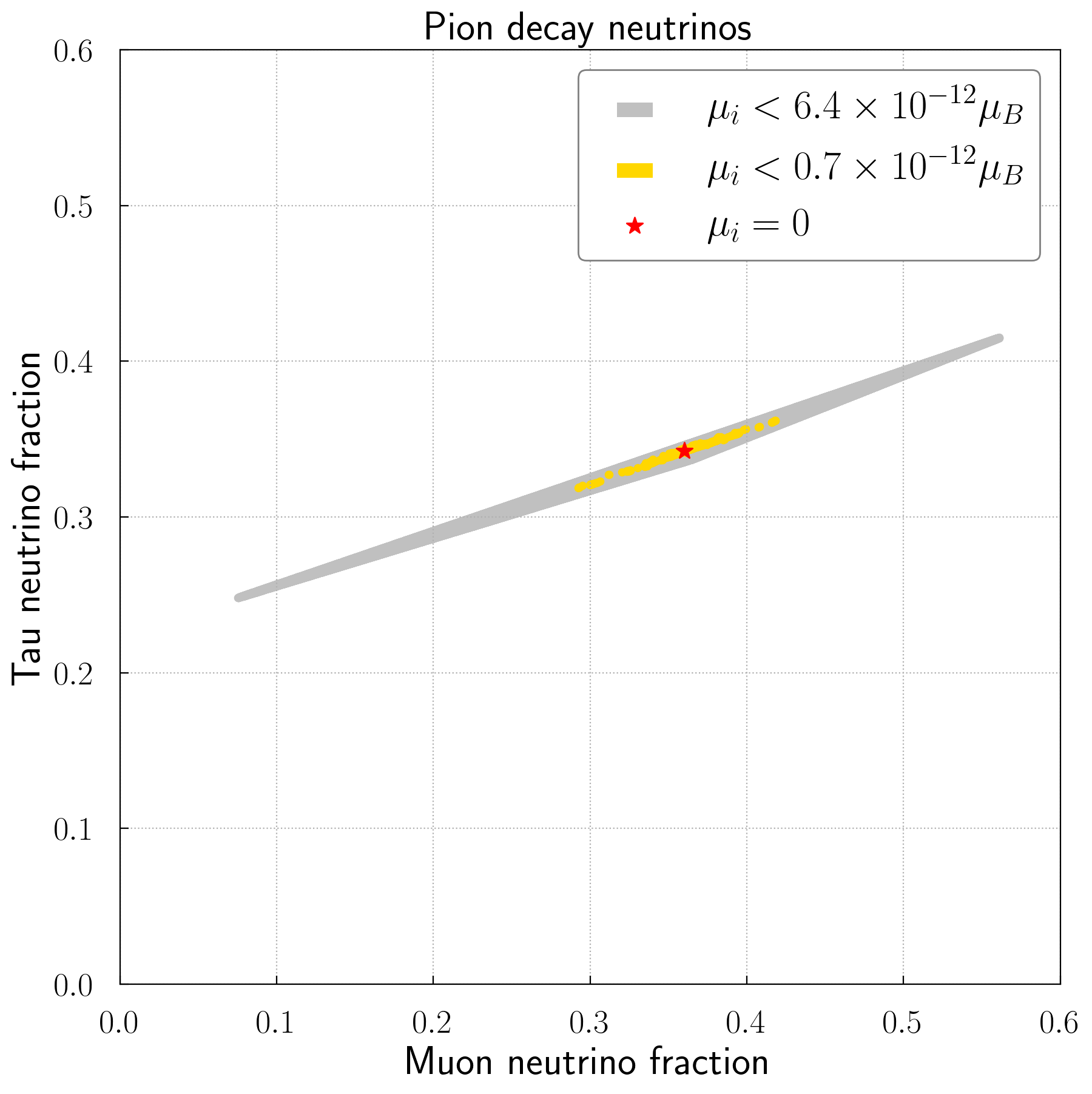}
	\caption{\label{fig:HE_fl_comp_1} Flavour compositions of high-energy neutrinos produced by $\pi^{\pm}$ decay after propagating from the Galactic centre for different ranges of neutrino magnetic moments.}
\end{figure}

In Figure \ref{fig:HE_fl_comp_1} we show the flavour compositions of high-energy neutrinos propagating in the Galactic magnetic field from the Galactic centre assuming $\pi^{\pm}$ decay neutrino production ($r^0$ = (1/3, 2/3, 0)). It is assumed that neutrino mixing parameters $\sin^2\theta_{12}$, $\sin^2\theta_{13}$, $\sin^2\theta_{23}$ and $\sin \delta_{CP}$ are given by the best-fit values from the \emph{NuFIT 5.1} global fit \cite{Esteban:2020cvm}. To obtain the flavour compositions we also vary neutrino magnetic moments $\mu_1$, $\mu_2$ and $\mu_3$ within the $10^{-13} \div 6.4\cdot10^{-12} \mu_B$ range. We conclude that interaction with the Galactic magnetic field indeed significantly modifies the flavour composition predicted by vacuum neutrino oscillations ($\mu_i = 0$) when the neutrino magnetic moments $\mu_i \sim 10^{-13} \mu_B$.

\begin{figure}[h]
	\centering 
	\includegraphics[width=.49\textwidth]{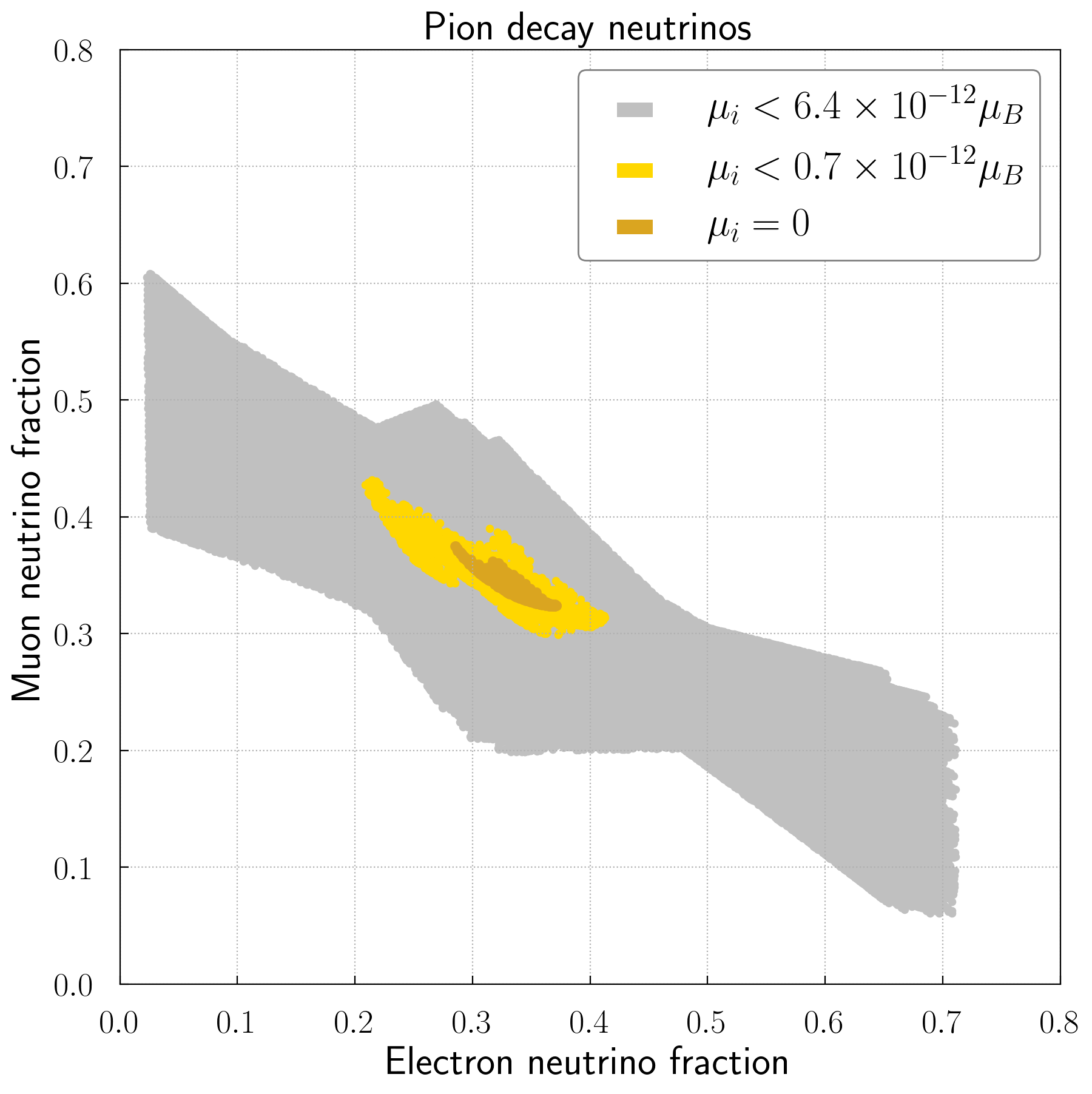}
	\includegraphics[width=.49\textwidth]{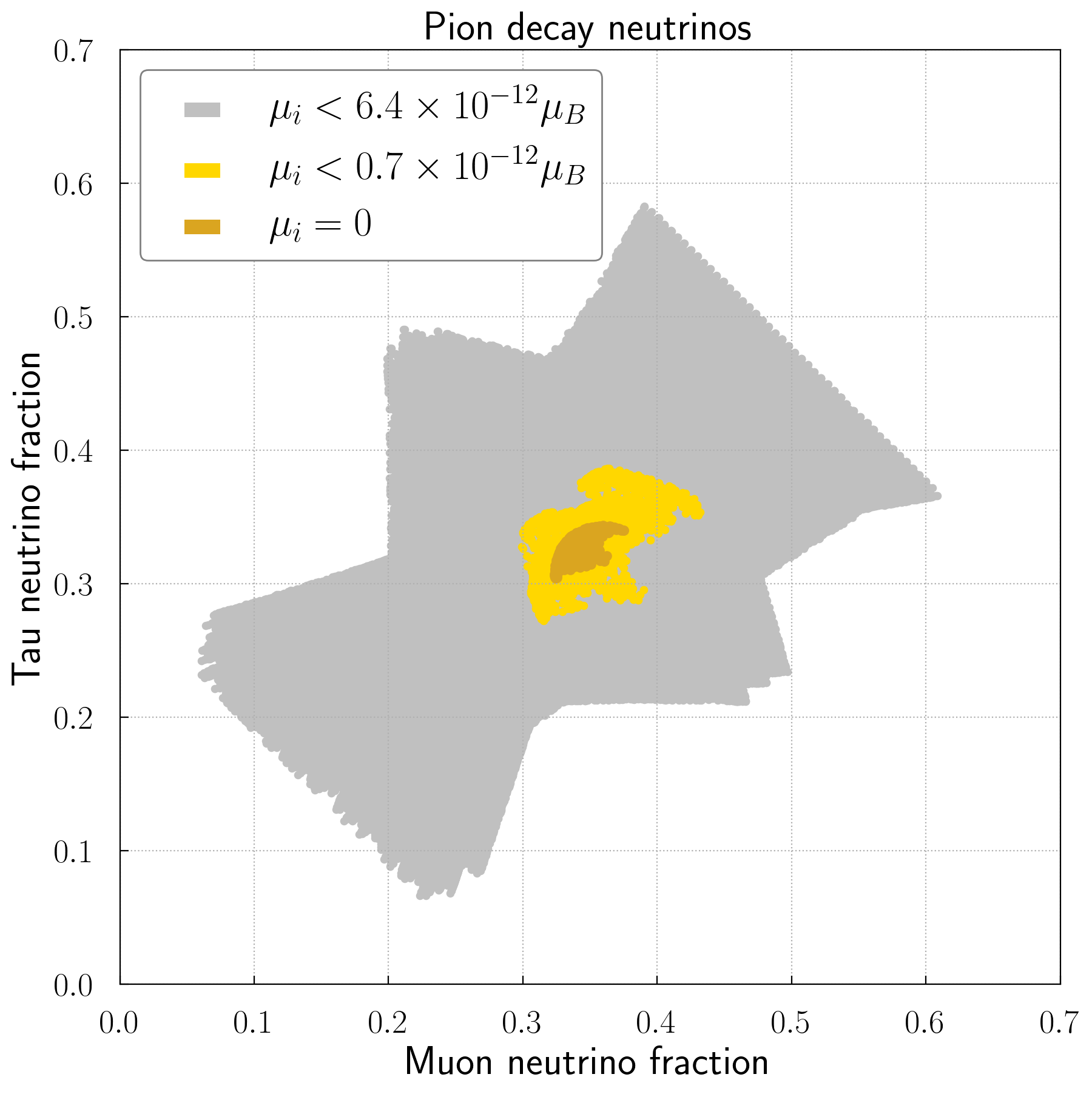}
	\caption{\label{fig:HE_fl_comp_2} Flavour compositions of high-energy neutrinos produced by $\pi^\pm$ decay after propagating from the Galactic centre for different ranges of neutrino magnetic moments accounting for current experimental errors of neutrino mixing parameters.}
\end{figure}

Presently, neutrino mixing parameters $\sin^2\theta_{12}$, $\sin^2\theta_{13}$, $\sin^2\theta_{23}$ and $\sin \delta_{CP}$ are measured with significant experimental errors. The predicted flavour compositions of high-energy neutrinos measured by the neutrino telescopes substantially depend on these errors \cite{Bustamante:2015waa}. In Figure \ref{fig:HE_fl_comp_2} the flavour compositions accounting for $3\sigma$ uncertainties in neutrinos mixing parameters are shown. By comparing Figure \ref{fig:HE_fl_comp_1} and Figure \ref{fig:HE_fl_comp_2} one can see that the experimental uncertainties significantly modify the range of possible neutrino flavour compositions. Future measurements of the neutrino mixing parameters, in particular of the CP-violating phase, can reduce these uncertainties and thus enhance the sensitivity of the future neutrino telescopes to the neutrino magnetic moments.

We also considered the case of muon decay high-energy neutrino production (the initial flavour composition $r^0$ = (0, 1, 0)). The results are presented in Figure \ref{fig:HE_fl_comp_3} and \ref{fig:HE_fl_comp_4} (see Appendix A).

To compute the high-energy neutrino flavour compositions form Figures \ref{fig:HE_fl_comp_1} and \ref{fig:HE_fl_comp_2}, we solved the neutrino wave packet evolution equation (\ref{WP_eq_approximate}) for different values of the wave packet widths $\sigma_x$. Confirming the results of Section 4A, we find no significant effects on the high-energy neutrino (100 TeV to 10 PeV) flavour composition stemming from deviation from the plane waves picture.

\subsection{Flavour composition of ultra-high energy neutrinos}
In this section we numerically examine oscillations and coherence of ultra high-energy neutrinos in a magnetic field. As it was shown in Section IIIA by Eq. (\ref{I_ij}) and Figure \ref{fig:integral}, in this case oscillations on vacuum frequencies may significantly contribute to the flavour composition as a result of the interplay between the oscillations on the vacuum and on the magnetic frequencies.

\begin{figure}[tbp]
	\centering 
	\includegraphics[width=.49\textwidth]{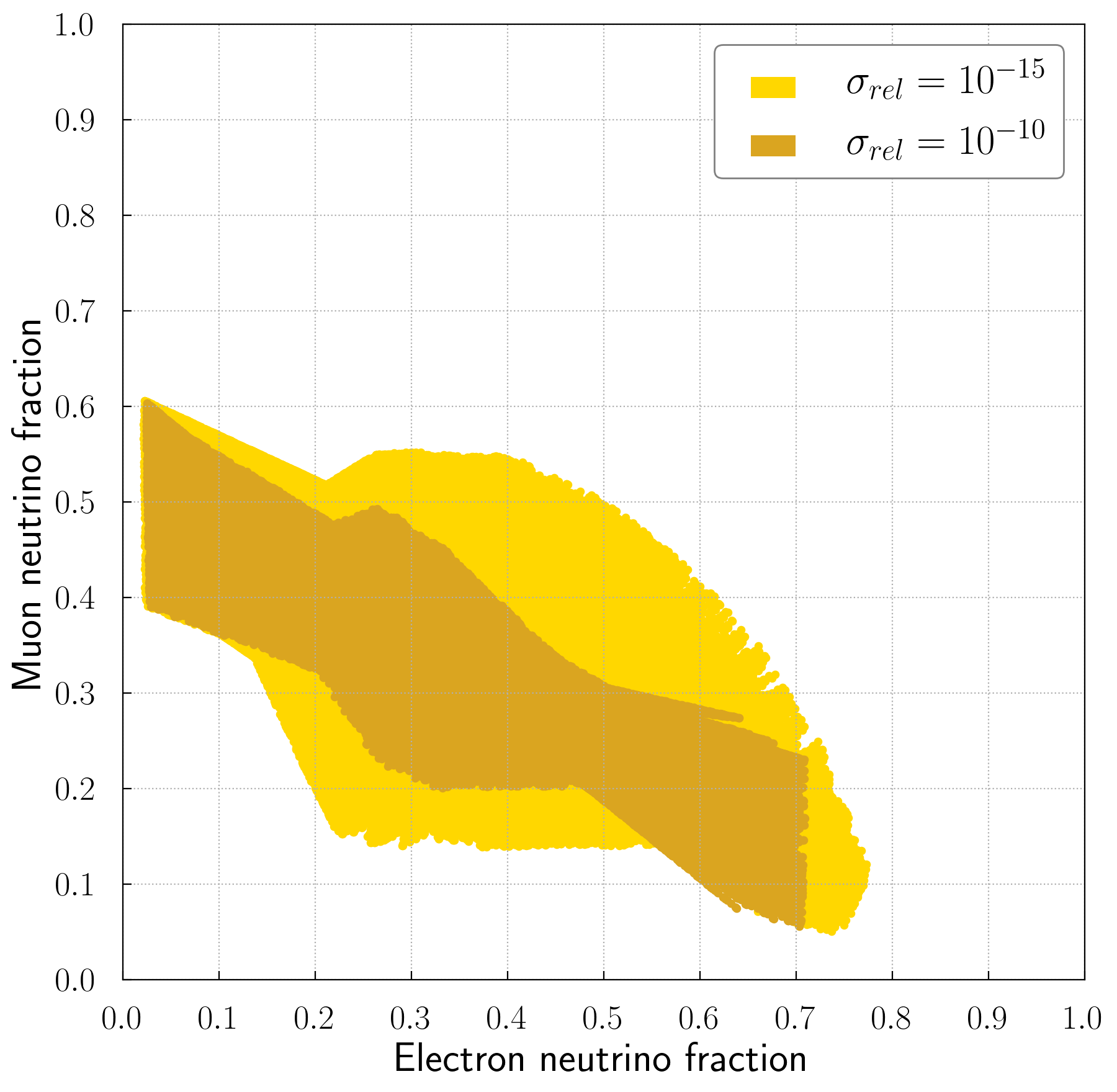}
	\includegraphics[width=.49\textwidth]{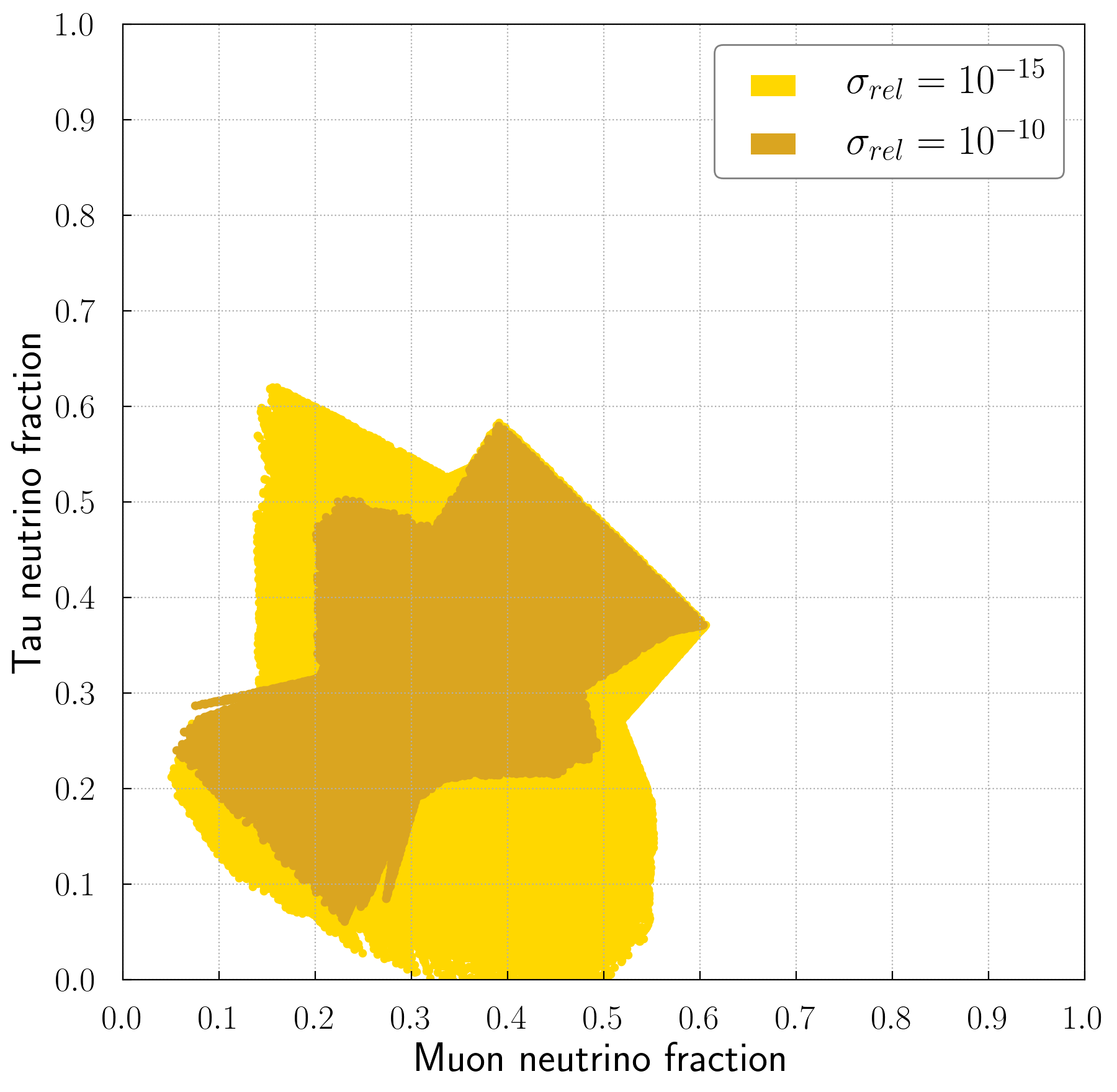}
	\caption{\label{fig:UHE_fl_comp_1} Flavour compositions of ultra-high energy neutrinos after propagating in the galactic magnetic field for different values of the wave packet parameter $\sigma_{rel} = \sigma_p/p_0$ and for spectral index $\gamma = 2$.}
\end{figure}

In Figure \ref{fig:UHE_fl_comp_1} we show possible flavour compositions of ultra-high energy neutrino for different values of the relative wave packet width $\sigma_{rel} = \sigma_p/p_0$ and different values of the neutrino magnetic moments $\mu_i < 10^{-13} \mu_B$. We use the same model of the magnetic field as in the previous section. We assume that the integration limits in (\ref{flux_integral}) are given by $E_{min} \approx 10^{21}$ eV and $E_{max} \approx 10^{22}$ eV. Previously, oscillations of neutrinos with such energies in a magnetic field were considered in \cite{Kurashvili:2017zab,Alok:2022pdn} without accounting for the wave packet separation effects. There are theoretically proposed mechanisms that can create neutrinos with even higher energies, for example cosmic strings are proposed as a source of neutrinos with energies up to $10^{25}$ eV \cite{Berezinsky_strings}. For a review of the possibility of the detection of ultra high-energy in future neutrino experiments see \cite{Ackermann:2022rqc}. Currently, there is no data on the ultra high-energy neutrino energy spectrum. We assume that it is described by the single power-law $d\Phi^0/dE \sim E^{-\gamma}$. To compute the flavour ratios shown in Figure \ref{fig:UHE_fl_comp_1}, we assumed that $\gamma = 2$. The possible flavour compositions for $\gamma = 2.5$ are shown in Figure \ref{fig:UHE_fl_comp_2} (see Appendix A).

For $\sigma_{rel} = 10^{-10}$ the flavour compositions from Figure \ref{fig:UHE_fl_comp_1} are similar to the those shown in Figure \ref{fig:HE_fl_comp_1}. In this case the vacuum coherence lengths $L_{coh}^{ijss}$ are smaller than the baseline $L$, and the neutrino massive states become fully decoherent. For $\sigma_{rel} = 10^{-15}$ the damping terms $D_{ij}^{vac} \approx 1$, and the additional distance-dependent contributions of the oscillations probabilities that are characterized by the the vacuum coherence lengths $L_{coh}^{ijss}$ are not suppressed. This leads to flavour compositions that significantly differ from the case of $\sigma_{rel} = 10^{-10}$. Thus, the ultra high-energy flavour composition indeed depends on the neutrino wave packets widths. Note that this conclusion is valid not only for the case of neutrino oscillations in a magnetic field, but also for the vacuum neutrino oscillations, since in the vacuum limit the expression for the flavour ratios (\ref{ratios_final}) also depends on $\tilde{C}_{ij}$ and $\tilde{S}_{ij}$ that are given by (\ref{c}) and (\ref{s}) and do not depend on the magnetic field strength.

So far, we only considered the flavour composition of the integrated neutrino fluxes $\Phi_{\beta}(L)$. Let us also briefly consider the flavour composition of the differential neutrino flux $\dfrac{d\Phi_{\beta}(L,E)}{dE}$. It depends on the probabilities of neutrino oscillations as follows

\begin{equation}
	\frac{d r_\alpha (L,E_{rec}) }{d E_{rec}} = \frac{d r_\alpha (L,E_{rec}) }{d E_{rec}}/\sum_\alpha \frac{d r_\alpha (L,E_{rec}) }{d E_{rec}} \sim \int P_{\alpha\beta}(L,E_{rec}) e^{-\frac{(E-E_{rec})^2}{2\delta_E^2 E_{rec}^2}} dE,
\end{equation}
where $E_{rec}$ is a reconstructed neutrino energy and $\delta_E$ is an energy reconstruction uncertainty. In the current neutrino telescopes for tracks $\delta_E \approx 0.15$, while for cascades $\delta_E \approx 2$.

The oscillations on vacuum frequencies are not suppressed when the following condition is satisfied:
\begin{equation}
\Delta \varphi (L,E) = \frac{\Delta m^2 L}{2E} - \frac{\Delta m^2 L}{2E(1+\delta_E)} \ll 2\pi.
\end{equation}

Assuming $\delta_E = 0.15$, we conclude that the oscillations on the vacuum frequencies and, consequently, the wave packet effects in the flavour composition $\dfrac{d r_\alpha (L,E_{rec}) }{d E_{rec}}$ are not suppressed for neutrino energies $E \le 10^{19}$ eV in the case of propagation in the Galactic magnetic field ($L \sim 10$ kpc) and for neutrino energies $E \le 10^{21}$ eV in the case of propagation in the extragalactic magnetic fields ($L \sim 1$ Mpc). The wave packet effects may become more pronounced, if the neutrino energy is measured with a lower uncertainty than $\delta_E = 0.15$.

Thus, the effects arising due to finite width of neutrino wave packet are indeed may be important in future for determining the flavour composition of ultra high-energy neutrinos.

\section{Conclusion}
\label{sec:conclusion}

In this paper the neutrino oscillations in a magnetic field are considered within the wave packet formalism, that provides the possibility to account for the effects of decoherence due to wave packet separation.

Within the developed formalism the expressions for the probabilities of the neutrino flavour and spin oscillations in a uniform magnetic field are obtained. It is shown that there are two different coherence lengths for oscillations on frequencies that depend on the magnetic field strength. The first one coincides with the coherence length for neutrino oscillations in vacuum, while the second one is proportional to $p_0^3$, and is significantly larger than the first coherence length for the case of high-energy neutrinos.

The evolution equation for the neutrino wave packet in a non-uniform magnetic field is derived. The numerical solution of this equation is obtained for the case of the high-energy neutrino propagation in the Galactic and extragalactic magnetic field. To describe the magnetic field of our Galaxy we use the model from \cite{Jansson:2012pc}.

The flavour compositions of high-energy neutrinos observed by a terrestrial telescope accounting for the effect of oscillations in the Galactic magnetic field are calculated. We assume that high-energy neutrinos are produced by the Galactic centre. Different high-energy neutrinos production mechanisms, namely $\pi^{\pm}$ decay and $\mu^{\pm}$  decay, are considered. It is shown that for the Dirac neutrinos interaction with a magnetic field can significantly modify the flavour composition if neutrino magnetic moments are $\sim 10^{-13}\mu_B$ and higher. While computing the flavour compositions, we account for experimental uncertainties in neutrino mixing parameters and magnetic field model parameters.

Decoherence of neutrino oscillations in a magnetic field due to wave packet separation is considered. It is shown that for the case of high-energy neutrinos from a galactic source partial decoherence, i.e. suppression of certain terms of the oscillations probabilities, occurs.

We also study decoherence of ultra-high energy neutrinos. In this case decoherence may not occur for certain values of the wave packets width. The impact of the decoherence effects on the neutrino flavour composition is analyzed.

In literature there are studies of the flavour ratios in the neutrino fluxes from different astrophysical sources accounting for the presence of a magnetic field (see, for instance, \cite{Kopp:2022cug} and references therein). In our previous paper \cite{Popov:2019nkr} the neutrino flavour, spin and spin-favour oscillations probabilities are calculated accounting for the whole set of possible conversions between four neutrino states. The obtained expressions for the neutrino oscillations probabilities exhibit new inherent features in the oscillation patterns. In particular, it is shown that in the presence of the transversal magnetic field for a given choice of parameters (the energy and magnetic moments of neutrinos and the strength of the magnetic field) the amplitude of the flavour oscillations
at the vacuum frequency is modulated by the magnetic field frequency. This phenomenon is important for evaluation of the neutrino flavour ratios and has never been considered elsewhere.

Note also that in our present study, as well as in \cite{Popov:2019nkr}, we account for the diagonal magnetic moments of three neutrino mass states. For the future, it would be important to combine the approaches to calculating neutrino ratios implemented in \cite{Kopp:2022cug} and the present study, which is based on the results of \cite{Popov:2019nkr}.

Note that the IceCube signal in the track channel from a source located in the southern sky, such as the Galactic center, is significantly contaminated by the atmospheric muons background. Thus, the Galactic centre neutrino emission is currently probed with the cascade channel that has higher angular uncertainty. The deployment of the neutrino telescopes in the Northern Hemisphere, such as Baikal-GVD and KM3NeT, will make the Galactic centre neutrino emission probes accessible via both the track and cascade channels. In \cite{Celli:2016uon} the authors estimate that the neutrino telescope, such as KM3NeT and Baikal-GVD, will be able to accumulate several events a year.

In the present paper we have mainly considered a galactic source of high-energy neutrinos, namely the Galactic centre. The developed theoretical formalism allows to describe neutrinos from extragalactic sources. Since magnetic fields of strengths $\mathcal{O}(\mu\text{G})$ are observed on Mpc scale \cite{Valee:2002}, for extragalactic neutrinos the effects of interaction with the magnetic field can be observable for neutrino magnetic moments $\sim 10^{-15}\mu_B$. The results of this paper can be also applied to describe supernova neutrino fluxes which in the future could be measured by JUNO, Hyper-Kamiokande and DUNE (see also in \cite{Kopp:2022cug}). The results are also of interest for description of neutrino flavour and spin oscillation inside astrophysical environments with strong magnetic fields, e.g. supernovae.


\acknowledgments
The work is supported by the Russian Science Foundation under grant No.24-12-00084.

\newpage
\section*{Appendix A}
\begin{figure}[h]
	\centering 
	\includegraphics[width=.49\textwidth]{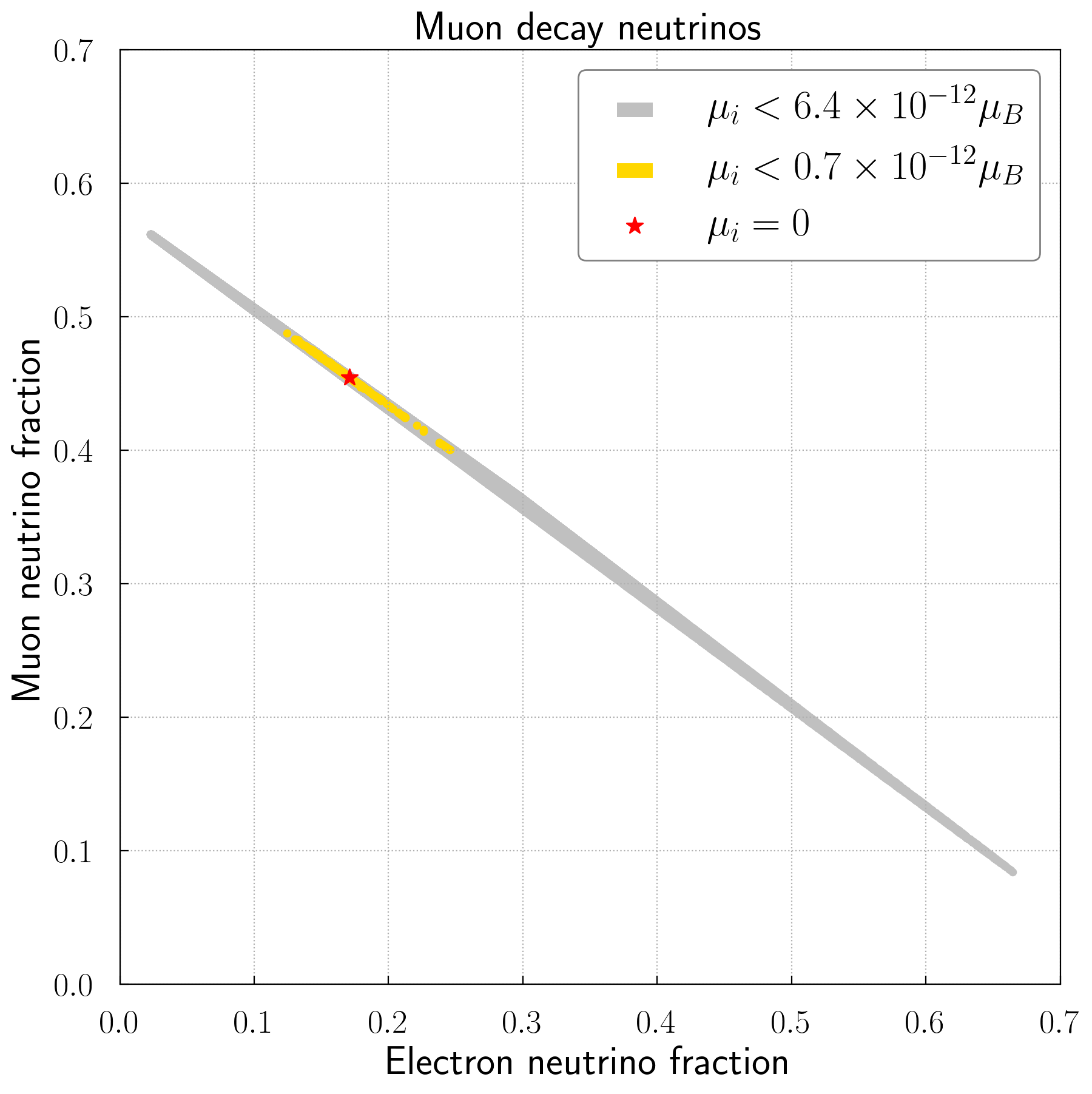}
	\includegraphics[width=.49\textwidth]{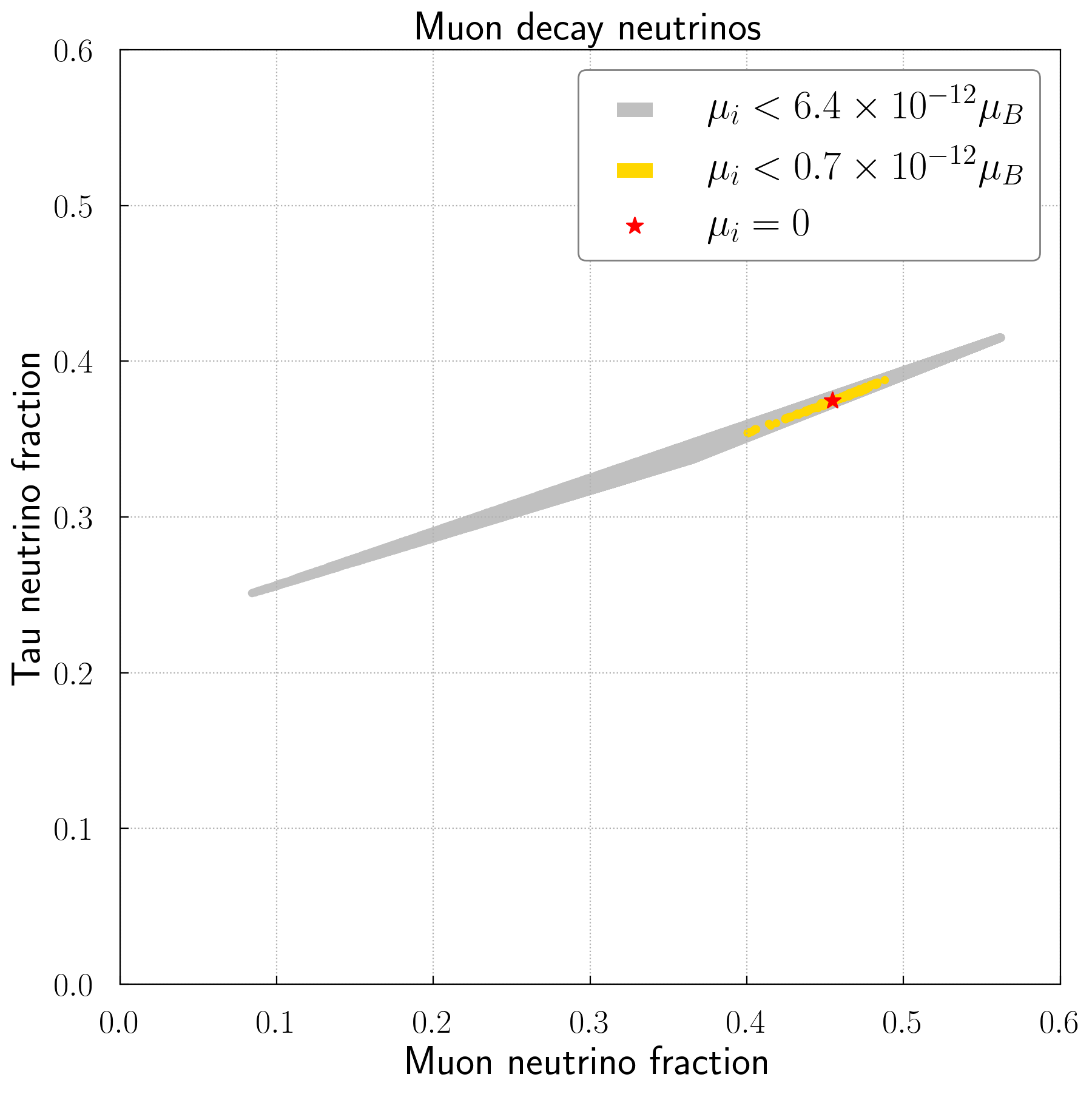}
	\caption{\label{fig:HE_fl_comp_3} Flavour compositions of high-energy neutrinos produced by $\mu^{\pm}$ decay after propagating from the Galactic centre for different ranges of the neutrino magnetic moments.}
\end{figure}
\begin{figure}[!]
	\centering 
	\includegraphics[width=.49\textwidth]{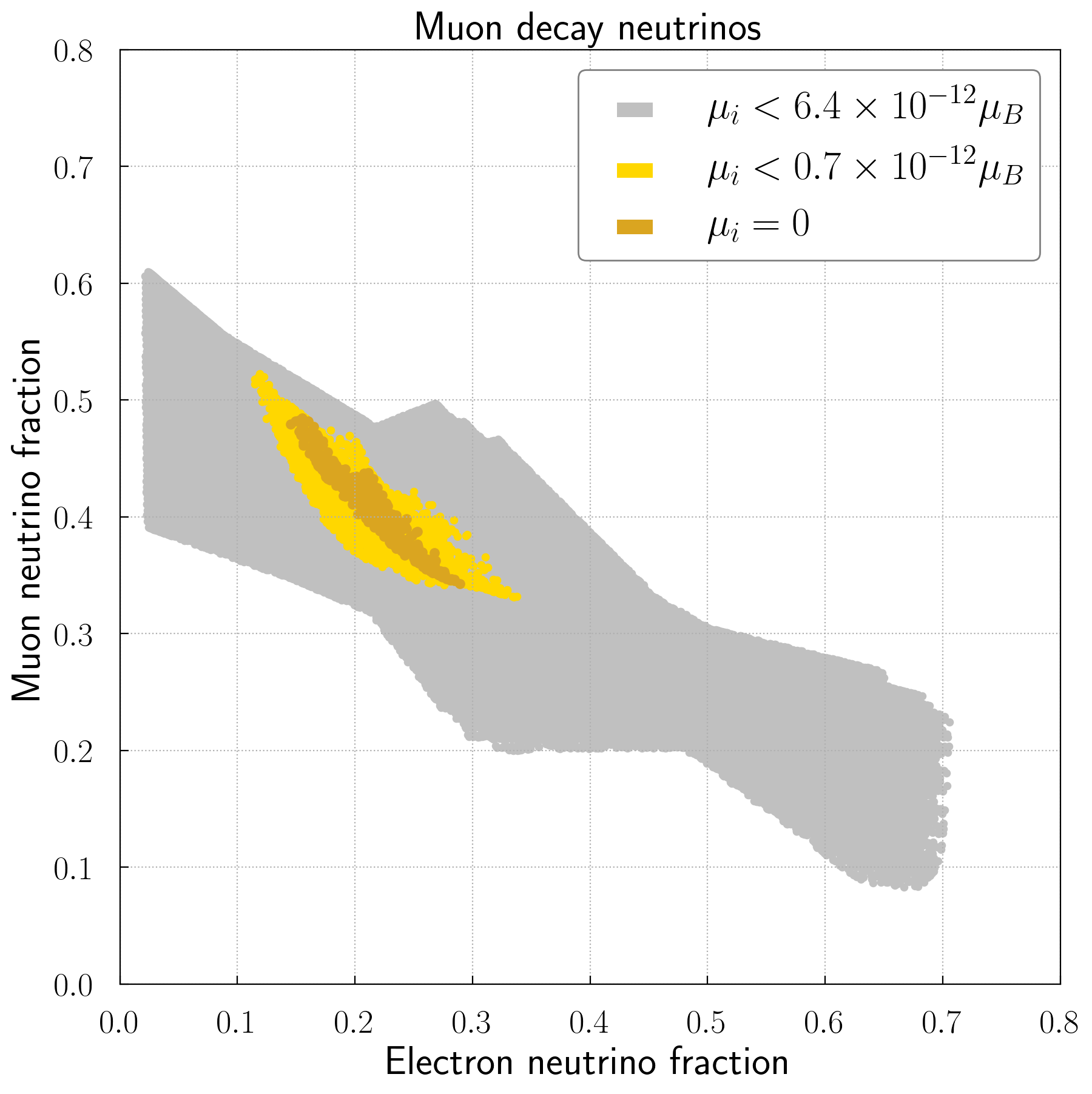}
	\includegraphics[width=.49\textwidth]{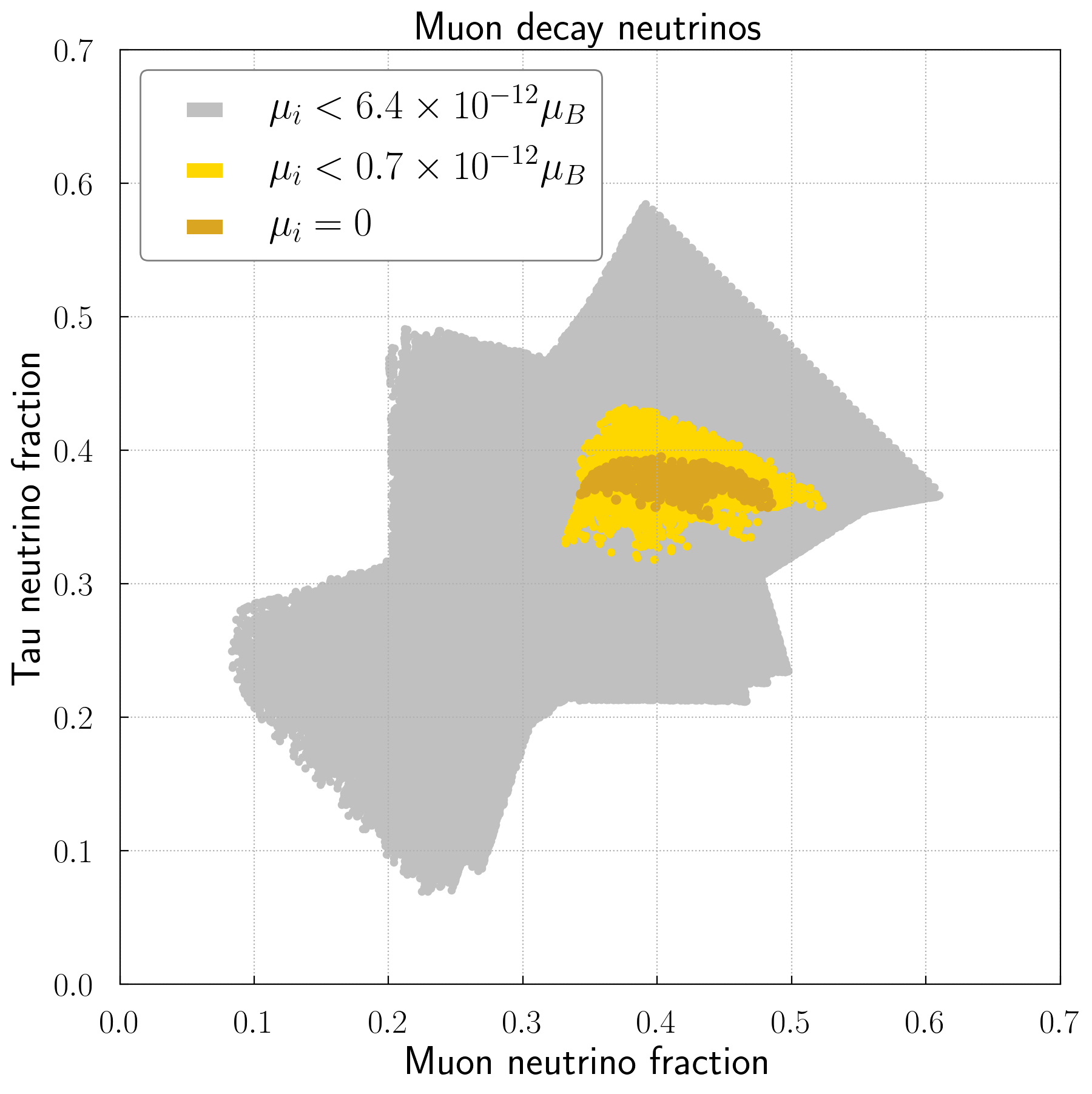}
	\caption{\label{fig:HE_fl_comp_4} Flavour compositions of high-energy neutrinos produced by $\mu^\pm$ decay after propagating from the Galactic centre for different ranges of neutrino magnetic moments accounting for current experimental errors of neutrino mixing parameters.}
\end{figure}

\begin{figure}[t]
	\centering 
	\includegraphics[width=.49\textwidth]{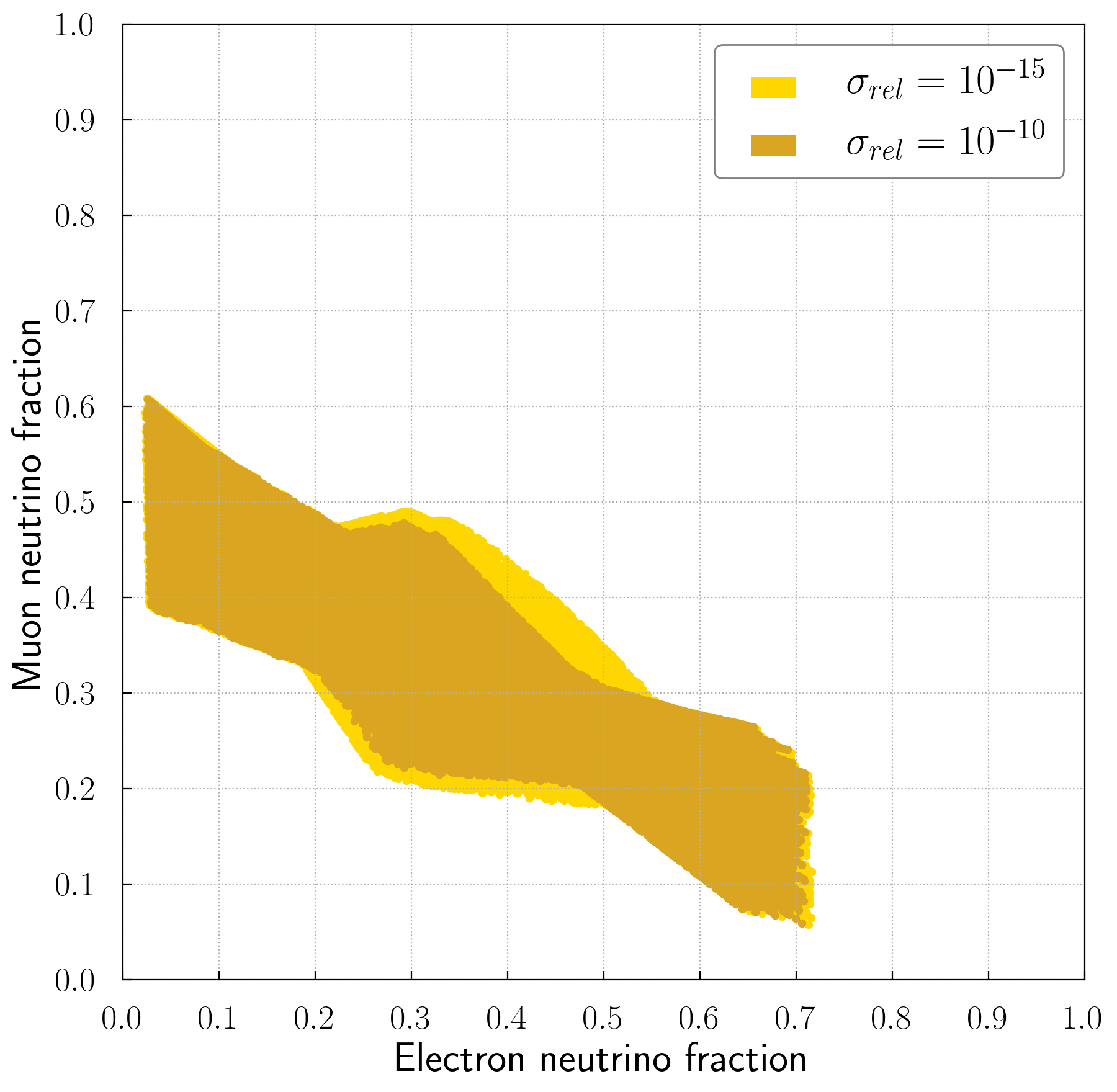}
	\includegraphics[width=.49\textwidth]{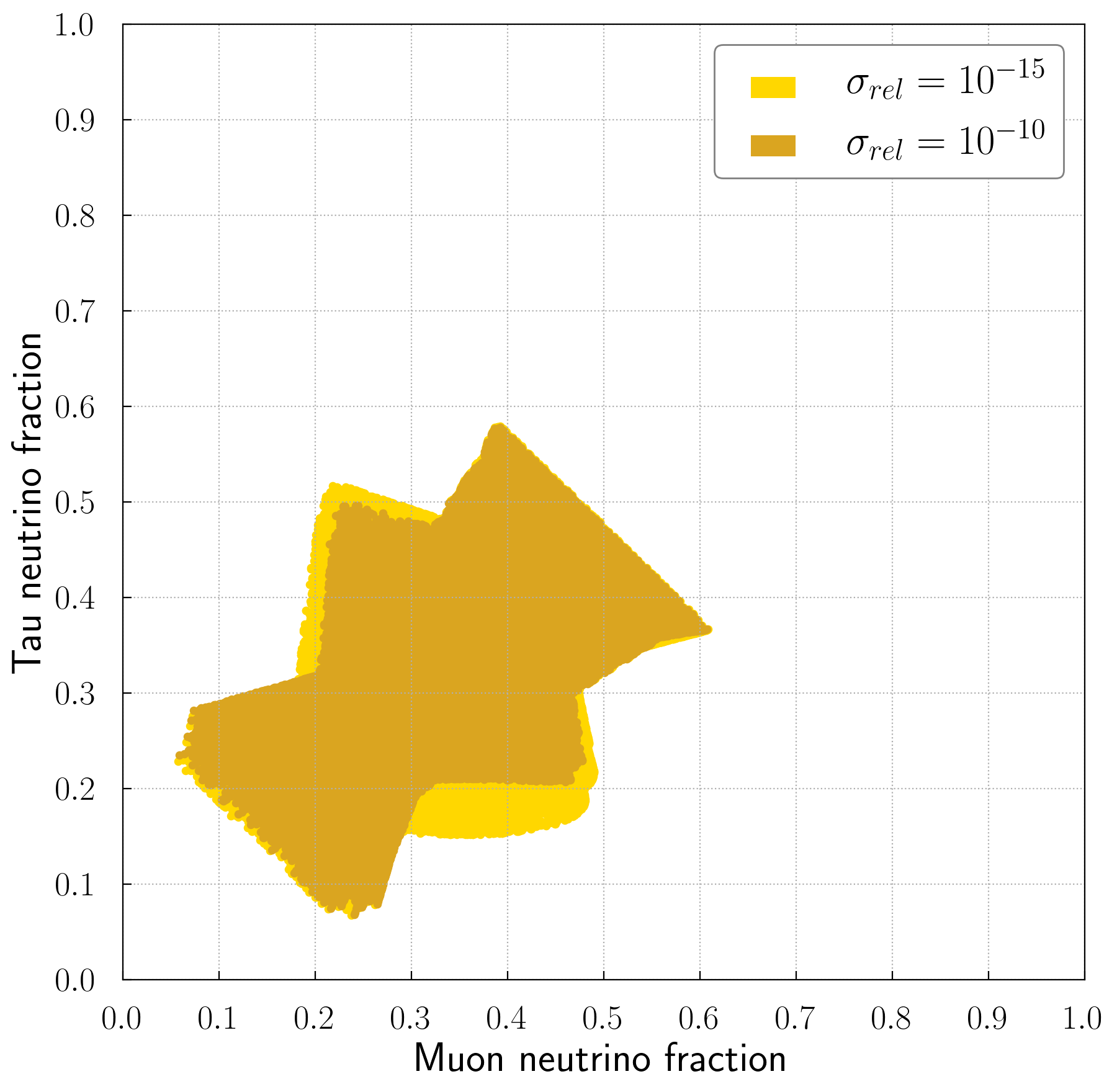}
	\caption{\label{fig:UHE_fl_comp_2} Flavour compositions of ultra-high energy neutrinos after propagating in the galactic magnetic field for different values of the wave packet parameter $\sigma_{rel} = \sigma_p/p_0$ and for spectral index $\gamma = 2.5$.}
\end{figure}

\end{document}